\newcommand{\CLASSINPUTtoptextmargin}{2cm}
\documentclass[journal,12pt,onecolumn,draftclsnofoot]{IEEEtran}
\renewcommand{\baselinestretch}{1.6}

\usepackage{cite}

\usepackage{textcomp}

\usepackage{cals}

\ifCLASSINFOpdf
 \usepackage[pdftex]{graphicx}
  \usepackage[labelsep=period]{caption}
 \usepackage{lipsum}
\usepackage{float}
\else
\fi

\usepackage{mathtools}  
\usepackage{amsmath}
\usepackage{amssymb,mathtools}
\usepackage{optidef}
\usepackage{bbm}
\usepackage{nicefrac}
\usepackage{tabulary}
\usepackage{booktabs}
\usepackage{color}
 \usepackage[export]{adjustbox}
\usepackage{blindtext}
\usepackage{multicol}
\usepackage{caption}
\usepackage{subcaption}
\usepackage{capt-of}
\usepackage[11pt]{moresize}
\usepackage{anyfontsize}
\DeclarePairedDelimiter\abs{\lvert}{\rvert}%
\DeclarePairedDelimiter\norm\lVert\rVert

\DeclareMathOperator*{\argmax}{arg\,max}

\usepackage{hhline}
\usepackage{comment}
\usepackage{caption}
\usepackage{subcaption}
\usepackage[font=small]{caption}
\usepackage{algorithm}
\usepackage{algorithmic}

\usepackage{setspace}
\usepackage{url}
\usepackage{enumitem}
\usepackage{booktabs}
\usepackage{multirow}
\makeatletter

\newcommand{\rmnum}[1]{\romannumeral #1}
\newcommand{\Rmnum}[1]{\expandafter\@slowromancap\romannumeral #1@}
\newcommand*{\rom}[1]{\expandafter\@slowromancap\romannumeral #1@}

\usepackage[utf8]{inputenc}
\usepackage[english]{babel}
\usepackage{amsthm}
 
\theoremstyle{definition}
\newtheorem{definition}{Definition}

\newtheorem{corollary}{Corollary}
\newtheorem{lemma}{Lemma}
\newtheorem{proposition}{Proposition}

\DeclarePairedDelimiterX{\expectarg}[1]{[}{]}{%
  \ifnum\currentgrouptype=16 \else\begingroup\fi
  \activatebar#1
  \ifnum\currentgrouptype=16 \else\endgroup\fi
}

\newcommand{\innermid}{\nonscript\;\delimsize\vert\nonscript\;}
\newcommand{\activatebar}{%
  \begingroup\lccode`\~=`\|
  \lowercase{\endgroup\let~}\innermid 
  \mathcode`|=\string"8000
}

\DeclareMathAlphabet{\mathcal}{OMS}{cmsy}{m}{n}
\SetMathAlphabet{\mathcal}{bold}{OMS}{cmsy}{b}{n}

\begin{document}

\title{Reinforcement Learning for Self Organization and Power Control of Two-Tier Heterogeneous Networks}

  \author{Roohollah~Amiri,~\IEEEmembership{Student Member,~IEEE,}
      Mojtaba~Ahmadi~Almasi,~\IEEEmembership{Student Member,~IEEE,}
      Jeffrey~G.~Andrews,~\IEEEmembership{Fellow,~IEEE,}
	  Hani~Mehrpouyan,~\IEEEmembership{Member,~IEEE}.
\thanks{This work was presented in part at ICC 2018~\cite{art_Amiri_ICC}.}
 \thanks{R. Amiri, M. A. Almasi, and H. Mehrpouyan are with Department of Electrical and Computer Engineering, Boise State University, Boise, ID, USA (e-mail: roohollahamiri@u.boisestate.edu; mojtabaahmadialm@u.boisestate.edu; hanimehrpouyan@boisestate.edu). J. G. Andrews (email:jandrews@ece.utexas.edu) is with the University of Texas at Austin, USA. Last revised \today.}}

\maketitle

\vspace{-35pt}
\begin{abstract}


Self-organizing networks (SONs) can help manage the severe interference in dense heterogeneous networks (HetNets). Given their need to automatically configure power and other settings, machine learning is a promising tool for data-driven decision making in SONs. In this paper, a HetNet is modeled as a dense two-tier network with conventional macrocells overlaid with denser small cells (e.g. femto or pico cells). First, a distributed framework based on multi-agent Markov decision process is proposed that models the power optimization problem in the network. Second, we present a systematic approach for designing a reward function based on the optimization problem. Third, we introduce Q-\textit{learning} based distributed power allocation algorithm (Q-DPA) as a self-organizing mechanism that enables ongoing transmit power adaptation as new small cells are added to the network. Further, the sample complexity of the Q-DPA algorithm to achieve $\epsilon$-optimality with high probability is provided. We demonstrate, at density of several thousands femtocells per km$^2$, the required quality of service of a macrocell user can be maintained via the proper selection of independent or cooperative learning and appropriate Markov state models. 
\end{abstract}
\vspace{-5pt}

%
\IEEEpeerreviewmaketitle

\section{Introduction}

Self-organization is a key feature as cellular networks densify and become more heterogeneous, through the additional small cells such as pico and femtocells \cite{art_self_0, art_self_1, art_what_5G, art_self_2, art_self_3}. Self-organizing networks (SONs) can perform self-configuration, self-optimization and self-healing. These operations can cover basic tasks such as configuration of a newly installed base station (BS), resource management, and fault management in the network~\cite{art_survey_SON}. In other words, SONs attempt to minimize human intervention where they use measurements from the network to minimize the cost of installation, configuration and maintenance of the network. In fact SONs bring two main factors in play: \textit{intelligence} and \textit{autonomous adaptability}~\cite{art_self_0, art_self_1}. Therefore, machine learning techniques can play a major role in processing underutilized sensory data to enhance the performance of SONs~\cite{art_bigdata_0, art_bigdata_1}. 

One of the main responsibilities of SONs is to configure the transmit power at various small BSs to manage interference. In fact, a small BS needs to configure its transmit power before joining the network (as self-configuration). Subsequently, it needs to dynamically control its transmit power during its operation in the network (as self-optimization). To address these two issues, we consider a macrocell network overlaid with small cells and focus on autonomous distributed power control, which is a key element of self-organization since it improves network throughput~\cite{art_powerControl,art_classic2,art_femto_2, art_Amiri_GSMM, art_nasir} and minimizes energy usage~\cite{art_PA_legacy,art_green,art_hasan}. We rely on local measurements, such as signal-to-interference-plus-noise ratio (SINR), and the use of machine learning to develop a SON framework that can continually improve the above performance metrics.


\subsection{Related Work}\label{sec_litSurvey}

In wireless communications, dynamic power control with the use of machine learning has been implemented via reinforcement learning (RL). In this context, RL is an area of machine learning that attempts to optimize a BS's transmit power to achieve a certain goal such as throughput maximization. One of the main advantages of RL with respect to supervised learning methods is its training phase, in which there is no need for correct input/output data. In fact, RL operates by applying the experience that it has gained through interacting with the network~\cite{book_sutton}. RL methods have been applied in the field of wireless communications in areas such as resource management~\cite{art_reward1, art_reward2, art_reward, art_bennis, art_femto, art_QL_ISJ}, energy harvesting~\cite{art_harvesting}, and opportunistic spectrum access~\cite{art_cognitive,art_spectrum_0}. A comprehensive review of RL applications in wireless communications can be found in~\cite{art_RL_context_awareness}.

Q-\textit{learning} is a model-free RL method~\cite{Watkins1992}. The model-free feature of Q-\textit{learning} makes it a proper method for scenarios in which the statistics of the network continuously change. Further, Q-\textit{learning} has low computational complexity and can be implemented by BSs in a distributed manner~\cite{art_Amiri_ICC}. Therefore, Q-\textit{learning} can bring \textit{scalability}, \textit{robustness}, and \textit{computational efficiency} to large networks. However, designing a proper reward function which accelerates the learning process and avoids false learning or unlearning phenomena~\cite{art_RW_0} is not trivial. Therefore, to solve an optimization problem, an appropriate reward function for Q-\textit{learning} needs to be determined.


In this regard, the works in~\cite{art_reward1, art_reward2, art_reward, art_bennis, art_femto, art_QL_ISJ} have proposed different reward functions to optimize power allocation between femtocell base stations (FBSs). The method in~\cite{art_reward1} uses independent Q-\textit{learning} in a cognitive radio system to set the transmit power of secondary BSs in a digital television system. The solution in~\cite{art_reward1} ensures that the minimum quality of service (QoS) for the primary user is met by applying Q-\textit{learning} and using the SINR as a metric. However, the approach in~\cite{art_reward1} doesn't take the QoS of the secondary users into considerations. The work in~\cite{art_reward2} uses cooperative Q-\textit{learning} to maximize the sum transmission rate of the femtocell users while keeping the transmission rate of macrocell users near a certain threshold. 
Further, the authors in~\cite{art_reward} have used the proximity of FBSs to a macrocell user as a factor in the reward function. This results in a fair power allocation scheme in the network. Their proposed reward function keeps the transmission rate of the macrocell user above a certain threshold while maximizing the sum transmission rate of FBSs. However, by not considering a minimum threshold for the FBSs' rates, the approach in~\cite{art_reward} fails to support some FBSs as the density of the network (and consequently interference) increases.
 The authors in~\cite{art_bennis} model the cross-tier interference management problem as a non-cooperative game between femtocells and the macrocell. In~\cite{art_bennis}, femtocells use the average SINR measurement to enhance their individual performances while maintaining the QoS of the macrocell user. In~\cite{art_femto}, the authors attempt to improve the transmission rate of cell-edge users while keeping the fairness between the macrocell and the femtocell users by applying a round robin approach. The work in~\cite{art_QL_ISJ} minimizes power usage in a Long Term Evolution (LTE) enterprise femtocell network by applying an exponential reward function without the requirement to achieve fairness amongst the femtocells in the network.
 
In the above works, the reward functions do not apply to dense networks. That is to say, first, there is no minimum threshold for the achievable rate of the femtocells. Second, the reward functions are designed to limit the macrocell user rate to its required QoS and not more than that. This property encourages an FBS to use more power to increase its own rate by assuming that the caused interference just affects the macrocell user. However, the neighbor femtocells suffer from this decision and overall the sum rate of the network decreases. Further, they do not provide a generalized framework for modeling a HetNet as a multi-agent RL network or a procedure to design a reward function which meets the QoS requirements of the network. In this paper, we focus on dense networks and try to provide a general solution to the above challenges.




\subsection{Contributions}

We propose a learning framework based on multi-agent Markov decision process (MDP). By considering an FBS as an agent, the proposed framework enables FBSs to join and adapt to a dense network autonomously. Due to unplanned and dense deployment of femtocells, providing the required QoS to all the users in the network becomes an important issue. Therefore, we design a reward function that trains the FBSs to achieve this goal. Furthermore, we introduce a Q-\textit{learning} based distributed power allocation approach (Q-DPA) as an application of the proposed framework. 
Q-DPA uses the proposed reward function to maximize the transmission rate of femtocells while prioritizing the QoS of the macrocell user. More specifically the contributions of the paper can be summarized as:

\begin{enumerate}[leftmargin=*]
\item We propose a framework that is agnostic to the choice of learning method but also connects the required RL analogies to wireless communications. The proposed framework models a multi-agent network with a single MDP that contains the joint action of the all the agents as its action set. Next, we introduce MDP factorization methods to provide a distributed and scalable architecture for the proposed framework. The proposed framework is used to benchmark the performance of different learning rates, Markov state models, or reward functions in two-tier wireless networks.

\item We present a systematic approach for designing a reward function based on the optimization problem and the nature of RL. In fact, due to scarcity of resources in a dense network, we propose some properties for a reward function to maximize sum transmission rate of the network while considering minimum requirements of all users. The procedure is simple and general and the designed reward function is in the shape of low complexity polynomials. Further, the designed reward function results in increasing the achievable sum transmission rate of the network while consuming considerably less power compared to greedy based algorithms.
\item We propose Q-DPA as an application of the proposed framework to perform distributed power allocation in a dense femtocell network. Q-DPA uses the factorization method to derive independent and cooperative learning from the optimal solution. Q-DPA uses local signal measurements at the femtocells to train the FBSs in order to: (\rmnum{1}) maximize the transmission rate of femtocells, (\rmnum{2}) achieve minimum required QoS for all femtocell users with a high probability, and (\rmnum{3}) maintain the QoS of macrocell user in a densely deployed femtocell network. In addition, we determine the minimum number of samples that is required to achieve an $\epsilon$-optimal policy in Q-DPA as its sample complexity.

\item We introduce four different learning configurations based on different combinations of independent/cooperative learning and Markov state models. We conduct extensive simulations to quantify the effect of different learning configurations on the performance of the network. Simulations show that the proposed Q-DPA algorithm can decrease power usage and as a result reduce the interference to the macrocell user.

\end{enumerate}

 The paper is organized as follows. In Section~\ref{sec_systemModel}, the system model is presented. Section~\ref{sec_optProblem} introduces the optimization problem and presents the existing challenges in solving this problem. Section~\ref{sec_framework} presents the proposed learning framework which models a two-tier femtocell network with a multi-agent MDP. Section~\ref{sec_QDPA} presents the Q-DPA algorithm as an application of the proposed framework. Section~\ref{sec_sim} presents the simulation results while Section~\ref{sec_con} concludes the paper.

\textit{\textbf{Notation:}} Lower case, boldface lower case, and calligraphic symbols represent scalars, vectors, and sets, respectively. For a real-valued function $Q : \mathcal{Z} \rightarrow \mathbb{R}$, $\norm{Q}$ denotes the max norm, i.e., $\norm{Q}=\underset{z \in \mathcal{Z}}{max}~\abs{Q\left(z\right)}$. $\mathbb{E}_{x}\left[\cdot\right]$, $\mathbb{E}_{x}\left[\cdot|\cdot\right]$, and $\frac{\partial f}{\partial x}$ denote the expectation, the conditional expectation, and the partial derivation with respect to $x$, respectively. Further, $\Pr\left(\cdot|\cdot\right)$ and $|\cdot|$ denote the conditional probability and absolute value operators, respectively.

\section{Downlink System Model}\label{sec_systemModel}
\begin{figure}
\begin{centering}
\includegraphics[width=.5\columnwidth]{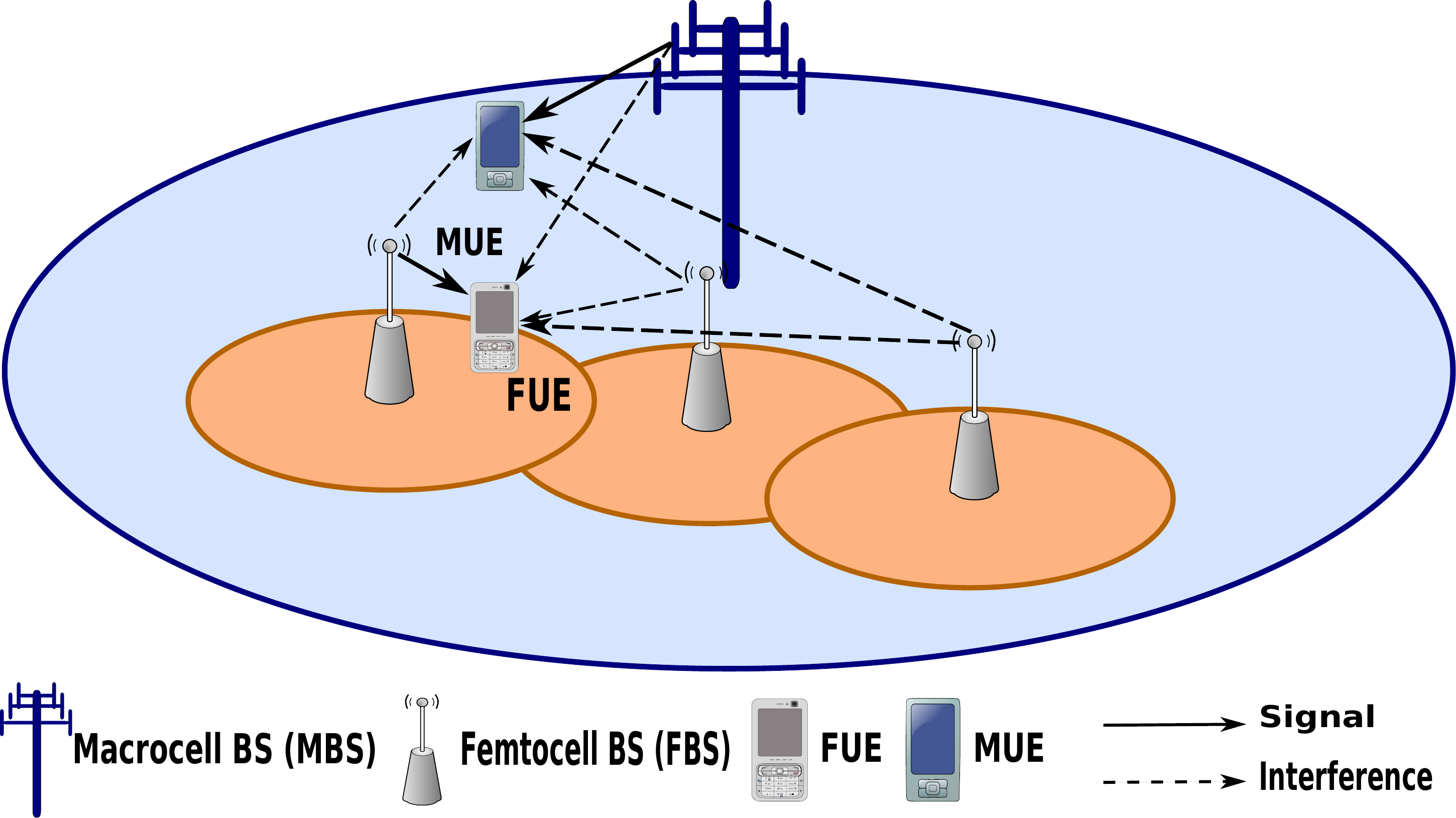}
\caption[width=.3\textwidth]{Macrocell and femtocells operating over the same frequency band.}
\label{fig_system}
\end{centering}
\end{figure}

Consider the downlink of a single cell of a HetNet operating over a set $\mathcal{S} = \left\lbrace 1, ..., S\right\rbrace$ of $S$ orthogonal subbands. In the cell a single macro base station (MBS) is deployed. The MBS serves one macrocell user equipment (MUE) over each subband while guaranteeing this user a minimum average SINR over each subband which is denoted by $\Gamma_0$. A set of FBSs are deployed in area of coverage of the macrocell. Each FBS selects a random subband and serves one femtocell user equipment (FUE). We assume that overall, on each subband $s \in \mathcal{S}$, a set $\mathcal{K}=\left\lbrace 1,...,K\right\rbrace$ of $K$ FBSs are operating. Each FBS guarantees a minimum average SINR denoted by $\Gamma_k$ to its related FUE. We consider a dense network in which the density results in both cross-tier and co-tier interference. Therefore, in order to control the interference-level and provide the users with their required minimum SINR, we focus on power allocation in the downlink of the femtocell network. Uplink results can be obtained in a similar fashion but are not included for brevity. The overall network configuration is presented in Fig.~\ref{fig_system}. We focus on one subband, meanwhile the proposed solution can be extended to a case in which each FBS supports multiple users on different subbands.

We denote the MBS-MUE pair by the index $0$ and the FBS-FUE pairs by the index $k$ from the set $\mathcal{K}$. In the downlink, the received signal at the MUE operating over subband $s$ includes interference from the femtocells and thermal noise. Hence, the SINR at the MUE operating over subband $s \in \mathcal{S}$, $\gamma_0$, is calculated as
\begin{align}\label{eq_sinr_mue}
\gamma_0=\frac{p_0 \lvert h_{0,0}\rvert^2 }{ \underbrace{\sum\limits_{k\in \mathcal{K}} p_k \lvert h_{k,0}\rvert^2}_{\text{femtocells' interference}} + N_0},
\end{align}
where $p_{0}$ denotes the power transmitted by the MBS and $h_{0,0}$ denotes the channel gain from the MBS to the MUE. Further, the power transmitted by the $k$th FBS is denoted by $p_k$ and the channel gain from the $k$th FBS to the MUE is denoted by $h_{k,0}$. Finally, $N_0$ denotes the variance of the additive white Gaussian noise. Similarly, the SINR at the $k$th FUE operating over subband $s \in \mathcal{S}$, $\gamma_k$, is obtained as
\vspace{-8pt}
\begin{equation}\label{eq_sinr_fue}
\gamma_k=\frac{p_k \lvert h_{k,k}\rvert^2 }{ \underbrace{p_0\lvert h_{0,k}\rvert^2}_{\text{macrocell's interference}} + \underbrace{\sum\limits_{j\in \mathcal{K} \setminus \left\lbrace k \right\rbrace} p_j \lvert h_{j,k}\rvert^2}_{\text{femtocells' interference}} + N_k},
\end{equation}
where $h_{k,k}$ denotes the channel gain between the $k$th FBS and the $k$th FUE, $h_{0,k}$ denotes the channel gain between the MBS and the $k$th FUE, $p_j$ denotes the transmit power of the $j$th FBS, $h_{j,k}$ is the channel gain between the $j$th FBS and the $k$th FUE, and $N_k$ is the variance of the additive white Gaussian noise. 
 Finally, the transmission rates, normalized by the transmission bandwidth, at the MUE and the FUE operating over subband $s \in \mathcal{S}$, i.e., $r_0$ and $r_k$, respectively, are expressed as $r_0 = \log_2 \left(1+\gamma_0\right)$ and $r_k= \log_2 \left(1+\gamma_k\right),~ k \in \mathcal{K}$.

\section{Problem Formulation}\label{sec_optProblem}
Each FBS has the objective of maximizing its transmission rate while ensuring that the SINR of the MUE is above the required threshold, i.e., $\Gamma_0$. Denoting $\mathbf{p}=\left\lbrace p_1,...,p_K\right\rbrace$ as the vector of the transmit powers of the $K$ FBSs operating over the subband $s \in \mathcal{S}$, the power allocation problem is presented as follow
\begin{maxi!}
{\mathbf{p}}{ \sum_{k \in \mathcal{K}} \log_2 \left(1+\gamma_k\right) \label{eq_objectiveopt}}{\label{eq_opt}}{}
\addConstraint{0}{\leq p_k}{\leq p_{max}}{\label{b}}{,~k \in \mathcal{K}}
\addConstraint{\gamma_0}{\geq \Gamma_0}{\label{c}}
\addConstraint{\gamma_k}{\geq \Gamma_k}{\label{d}}{,~k \in \mathcal{K}}.
\end{maxi!}
\noindent
where $p_{max}$ defines the maximum available transmit power at each FBS. The objective~\eqref{eq_objectiveopt} is to maximize the sum transmission rate of the FUEs. Constraint \eqref{b} refers to the power limitation of every FBS. Constraints~\eqref{c} and~\eqref{d} ensure that the minimum SINR requirement is satisfied for the MUE and the FUEs. The addition of constraint~\eqref{d} to the optimization problem is one of the differences between the proposed approach in this paper and that of~\cite{art_reward1, art_reward2, art_reward, art_bennis, art_femto, art_QL_ISJ}.

Considering \eqref{eq_sinr_fue}, it can be concluded that the optimization in~\eqref{eq_opt} is a non-convex problem for dense networks. This follows from the SINR expression in \eqref{eq_sinr_fue} and the objective function \eqref{eq_opt}. More specifically, the interference term due to the neighboring femtocells in the denominator of \eqref{eq_sinr_fue} ensures that the optimization problem in \eqref{eq_opt} is not convex~\cite{art_cvx_1}. This interference term may be ignored in low density networks but cannot be ignored in dense networks consisting of a large number of femtocells~\cite{art_slmz}. However, non-convextiy is not the only challenge of the above problem. In fact, many iterative algorithms are developed to solve the above optimization problem with excellent performance. However, their algorithms contains expensive computations such as matrix inversion and bisection or singular value decomposition in each iteration which makes their real-time implementation challenging~\cite{art_DNN}. Besides, the $k$th FBS is only aware of its own transmit power, $p_k$, and does not know the transmit powers of the remaining FBSs. Therefore, the idea here is to treat the given problem as a black-box and try to learn the relation between the transmit power and the resulting transmission rate gradually by interacting with the network and simple computations.


To realize self-organization, each FBS should be able to operate autonomously. This means an FBS should be able to connect to the network at anytime and to continuously adapt its transmit power to achieve its objectives. Therefore, our optimization problem requires a self-adaptive solution. The steps for achieving self-adaptation can be summarized as: (\rmnum{1}) the FBS measures the interference level at its related FUEs, (\rmnum{2}) determines the maximum transmit power to support its FUEs while not greatly degrading the performance of other users in the network. In the next section, the required framework to solve this problem will be presented.

\section{The Proposed Learning Framework}\label{sec_framework}
Here, first we model a multi-agent network as an MDP. Then the required definitions, evaluation methods, and factorization of the MDP to develop a distributed learning framework are explained. Subsequently, the femtocell network is modeled as a multi-agent MDP and the proposed learning framework is developed. 

\vspace{-5pt}
\subsection{Multi-Agent MDP and Policy Evaluation}\label{sec_MDP}

A single-agent MDP comprises an agent, an environment, an action set, and a state set. The agent can transition between different states by choosing different actions. The trace of actions that is taken by the agent is called its policy. With each transition, the agent will receive a reward from the environment, as a consequence of its action, and will save the discounted summation of rewards as a cumulative reward. The agent will continue its behavior with the goal of maximizing the cumulative reward and the value of cumulative reward evaluates the chosen policy. The discount property increases the impact of recent rewards and decreases the effect of later ones. If the number of transitions is limited, the non-discounted summation of rewards can be used as well.

A multi-agent MDP consists of a set, $\mathcal{K}$, of $K$ agents. The agents select actions to move between different states of the model to maximize the cumulative reward received by all the agents. Here, we again formulate the network of agents as one MDP, e.g., we define the action set as the joint action set of all the agents. Therefore, the multi-agent MDP framework is defined with a tuple as $\left(\mathcal{A}, \mathcal{X}, Pr, \mathbf{R} \right)$ with the following definitions.
\begin{itemize}[leftmargin=*]
\item  $\mathcal{A}$ is the joint set of all the agents' actions. An agent $k$ selects its action $a$ from its action set $\mathcal{A}_k$, i.e., $a_k \in \mathcal{A}_k$. The joint action set is represented as $\mathcal{A}= \mathcal{A}_1 \times \cdots \times \mathcal{A}_K$, with $\mathbf{a}\in \mathcal{A}$ as a single joint action.
\item The state of the system is defined with a set of random variables. Each random variable is represented by $X_i$ with $i=1,...,n$, and the state set is represented as $\mathcal{X}=\left\lbrace X_1,X_2,...,X_n\right\rbrace$, where $\mathbf{x} \in \mathcal{X}$ denotes a single state of the system. Each random variable reflects a specific feature of the network.
\item The transition probability function, $\Pr\left(\mathbf{x},\mathbf{a},\mathbf{x}'\right)$, represents the probability of taking joint action $\mathbf{a}$ at state $\mathbf{x}$ and ending in state $\mathbf{x}^\prime$. In other words, the transition probability function defines the environment which agents are interacting with. 
\item $\mathbf{R}\left(\mathbf{x},\mathbf{a}\right)$ is the reward function such that its value is the received reward by the agents for taking joint action $\mathbf{a}$ at state $\mathbf{x}$.
\end{itemize}

We define $\pi:\mathcal{X}\rightarrow\mathbf{A}$ as the policy function, where $\pi\left(\mathbf{x}\right)$ is the joint action that is taken at the state $\mathbf{x}$. In order to evaluate the policy $\pi\left(\mathbf{x}\right)$, a value function $V_{\pi}\left(\mathbf{x}\right)$ and an action-value function $\mathbf{Q}_{\pi}\left(\mathbf{x},\mathbf{a}\right)$ are defined. The value of the policy $\pi$ in state $\mathbf{x}^{\prime} \in \mathcal{X}$ is defined as~\cite{book_sutton}
\begin{align}\label{eq_value_func}
V_{\pi}\left(\mathbf{x}^{\prime}\right) = \mathbb{E}_{\pi} \left[ \sum_{t=0}^{\infty} \beta^t \mathbf{R}^{\left(t+1\right)}  \left\vert \mathbf{x}^{\left(0\right)} = \mathbf{x}^{\prime} \right. \right],
\end{align}
in which $\beta \in \left( 0,1\right]$ is a discount factor, $\mathbf{R}^{\left(t+1\right)}$ is the received reward at time step $t+1$, and $\mathbf{x}^{\left(0\right)}$ is the initial state. 
The action-value function, $\mathbf{Q}_{\pi}\left(\mathbf{x},\mathbf{a}\right)$, represents the value of the policy $\pi$ for taking joint action $\mathbf{a}$ at state $\mathbf{x}$ and then following policy $\pi$ for subsequent iterations. According to~\cite{book_sutton}, the relation between the value function and the action-value function is given by
\vspace{-10pt}
\begin{align}\label{eq_state_value_func}
\mathbf{Q}_{\pi}\left(\mathbf{x},\mathbf{a}\right) = \mathbf{R}\left(\mathbf{x},\mathbf{a}\right) + \beta \sum_{\mathbf{x}^\prime \in \mathcal{X}} \Pr\left(\mathbf{x}^\prime|\mathbf{x},\mathbf{a}\right) V_{\pi}\left(\mathbf{x}^\prime\right).
\end{align}
For the ease of notation, we will use $V$ and $\mathbf{Q}$ for the value function and the action-value function of policy $\pi$, respectively. Further, we use the term \textit{Q-function} to refer to the action-value function. The optimal value of state $\mathbf{x}$ is the maximum value that can be reached by following any policy and starting at this state. An optimal value function $V^*$, which gives an optimal policy $\pi^*$, satisfies the Bellman optimality equation as~\cite{book_sutton}
\begin{align}\label{eq_Bellman}
V^*\left(\mathbf{x}\right) = \underset{\mathbf{a}}\max~\mathbf{Q}^*\left(\mathbf{x},\mathbf{a}\right),
\end{align}
where $\mathbf{Q}^*\left(\mathbf{x},\mathbf{a}\right)$ is an optimal Q-function under policy $\pi^*$. 
 The general solution for~\eqref{eq_Bellman} is to start from an arbitrary policy and using the \textit{generalized policy iteration} (GPI)~\cite{book_sutton} method to iteratively evaluate and improve the chosen policy to achieve an optimal policy. If the agents have \textit{a priori} information of the environment, i.e., $Pr\left(\mathbf{x},\mathbf{a},\mathbf{x}'\right)$ is known to the agents, dynamic programming is the solution for~\eqref{eq_Bellman}. However, the environment is unknown in most practical applications. Hence, we rely on reinforcement learning (RL) to derive an optimal Q-function. 
 RL uses temporal-difference to provide a real-time solution for the GPI method~\cite{book_sutton}. As a result, in Section~\ref{sec_QDPA}, we use Q-\textit{learning}, as a specific method of RL, to solve~\eqref{eq_Bellman}.

\subsection{Factored MDP}\label{sec_factored}

To this point, we defined the Q-function over the joint state-action space of all the agents, i.e., $\mathcal{X} \times \mathcal{A}$. We refer to this Q-function as the global Q-function. According to~\cite{Watkins1992}, Q-\textit{learning} finds the optimal solution to a single MDP with probability one. However, in large MDPs, due to exponential increase in the size of the joint state-action space with respect to the number of agents, the solution to the problem becomes intractable. To resolve this issue, we use \textit{factored} MDPs as a decomposition technique for large MDPs. The idea in factored MDPs is that many large MDPs are generated by systems with many parts that are weakly interconnected. Each part has its associated state variables and the state space can be factored into subsets accordingly. The definition of the subsets affects the optimality of the solution~\cite{art_factor_MDP}, and investigating the optimal factorization method helps with understanding the optimality of multi-agent RL solutions~\cite{art_amiri_vtc}. In~\cite{art_factor_MDP_1} power control of a multi-hop network is modeled as an MDP and the state set is factorized into multiple subsets each referring to a single hop. The authors in~\cite{art_Guestrin} show that the subsets can be defined based on the local knowledge of the agents from the environment. Meanwhile, we aim to distribute the power control to the nodes of the network. Therefore, due to the definition of the problem in Section~\ref{sec_optProblem} and the fact that each FBS is only aware of its own power, we use the assumption in~\cite{art_Guestrin} and define the individual action set of the agents, i.e., $\mathcal{A}_k$, as the subsets of the joint action set. Consequently, the resultant Q-function for the $k$th agent is defined as $Q_k\left(\mathbf{x}_k,a_k \right)$, in which $a_k \in \mathcal{A}_k$, $\mathbf{x}_k \in \mathcal{X}_k$ is the state vector of the $k$th agent, and $\mathcal{X}_k,~k\in \mathcal{K}$, are the subsets of the global state set of the system, i.e., $\mathcal{X}$. 

In factored MDPs, We assume that the reward function is factored based on the subsets, i.e.,
\vspace{-5pt}
	\begin{align}\label{reward_ass}
	\mathbf{R}\left(\mathbf{x},\mathbf{a}\right) = \sum_{k \in \mathcal{K}} R_k\left(\mathbf{x}_k,a_k\right),
	\end{align}
	\vspace{-5pt}
where, $R_k\left(\mathbf{x}_k,a_k\right)$ is the local reward function of the $k$th agent. Moreover, we also assume that the transition probabilities are factored, i.e., for the $k$th subsystem we have
\vspace{-5pt}
\begin{align}\label{eq_fact_trans}
\Pr\left(\mathbf{x}^\prime_k|\mathbf{x},\mathbf{a}\right) = \Pr\left(\mathbf{x}^\prime_k|\mathbf{x}_k,a_k\right),~\left(\mathbf{x},\mathbf{a}\right)\in \mathcal{X}\times \mathcal{A},~\left(\mathbf{x}_k,a_k\right)\in \mathcal{X}_k \times \mathcal{A}_k,~\mathbf{x}^\prime_k \in \mathcal{X}_k.
\end{align}
\vspace{-5pt}
The value function for the global MDP is given by
	\begin{align}
\mathbf{V}\left(\mathbf{x}\right) = \mathbb{E}\left[ \sum_{t=0}^{\infty} \beta^t \mathbf{R}^{\left(t+1\right)}\left(\mathbf{x},\mathbf{a}\right) \right]=\mathbb{E} \left[ \sum_{t=0}^{\infty} \beta^t \sum_{k \in \mathcal{K}} R_k^{\left(t+1\right)}\left(\mathbf{x}_k,a_k\right) \right]=\sum_{k \in \mathcal{K}} V_k\left(\mathbf{x}_k\right),
	\end{align}
where, $V_k\left(\mathbf{x}_k\right)$ is the value function of the $k$th agent. Therefore, the derived policy has the value function equal to the linear combination of local value functions. Further, according to~\eqref{eq_state_value_func}, for each agent $k \in \mathcal{K}$
\vspace{-10pt}
	\begin{align}
Q_k\left(\mathbf{x}_k,a_k\right) = R_k\left(\mathbf{x}_k,a_k\right) + \beta \sum_{\mathbf{x}_k^\prime} \Pr\left(\mathbf{x}^\prime_k|\mathbf{x}_k,a_k\right) V_k\left(\mathbf{x}^\prime_k\right),
	\end{align}
	\vspace{-10pt}
and for the global Q-function
\begin{equation}\label{eq_approx}
	\begin{aligned}
\mathbf{Q}\left(\mathbf{x},\mathbf{a}\right) &= \mathbf{R}\left(\mathbf{x},\mathbf{a}\right) + \beta \sum_{\mathbf{x}^\prime \in \mathcal{X}} \Pr\left(\mathbf{x}^\prime|\mathbf{x},\mathbf{a}\right) \mathbf{V}\left(\mathbf{x}^\prime\right)  \\&
=\sum_{k \in \mathcal{K}} R_k\left(\mathbf{x}_k,a_k\right) + \beta \sum_{\mathbf{x}^\prime \in \mathcal{X}} \Pr\left(\mathbf{x}^\prime|\mathbf{x},\mathbf{a}\right) \sum_{k \in \mathcal{K}} V_k\left(\mathbf{x}_k\right)\\&
=\sum_{k \in \mathcal{K}} R_k\left(\mathbf{x}_k,a_k\right) + \beta \sum_{k \in \mathcal{K}} \sum_{\mathbf{x}^\prime_k \in \mathcal{X}_k} \Pr\left(\mathbf{x}^\prime_k|\mathbf{x},\mathbf{a}\right) V_k\left(\mathbf{x}_k\right)\\&
=\sum_{k \in \mathcal{K}} R_k\left(\mathbf{x}_k,a_k\right) + \beta \sum_{k \in \mathcal{K}} \sum_{\mathbf{x}^\prime \in \mathcal{X}_k} \Pr\left(\mathbf{x}^\prime_k|\mathbf{x}_k,a_k\right) V_k\left(\mathbf{x}_k\right)
=\sum_{k \in \mathcal{K}} Q_k\left(\mathbf{x}_k,a_k\right).
	\end{aligned}
\end{equation}
Therefore, based on the assumptions in~\eqref{reward_ass} and~\eqref{eq_fact_trans}, the global Q-function can be approximated with the linear combination of local Q-functions. Further,~\eqref{eq_approx} results in a distributed and scalable architecture for the framework.

\subsection{Femtocell Network as Multi-Agent MDP}\label{sec_femtocell_MDP}

\begin{figure}
\begin{centering}
\includegraphics[width=0.4\columnwidth]{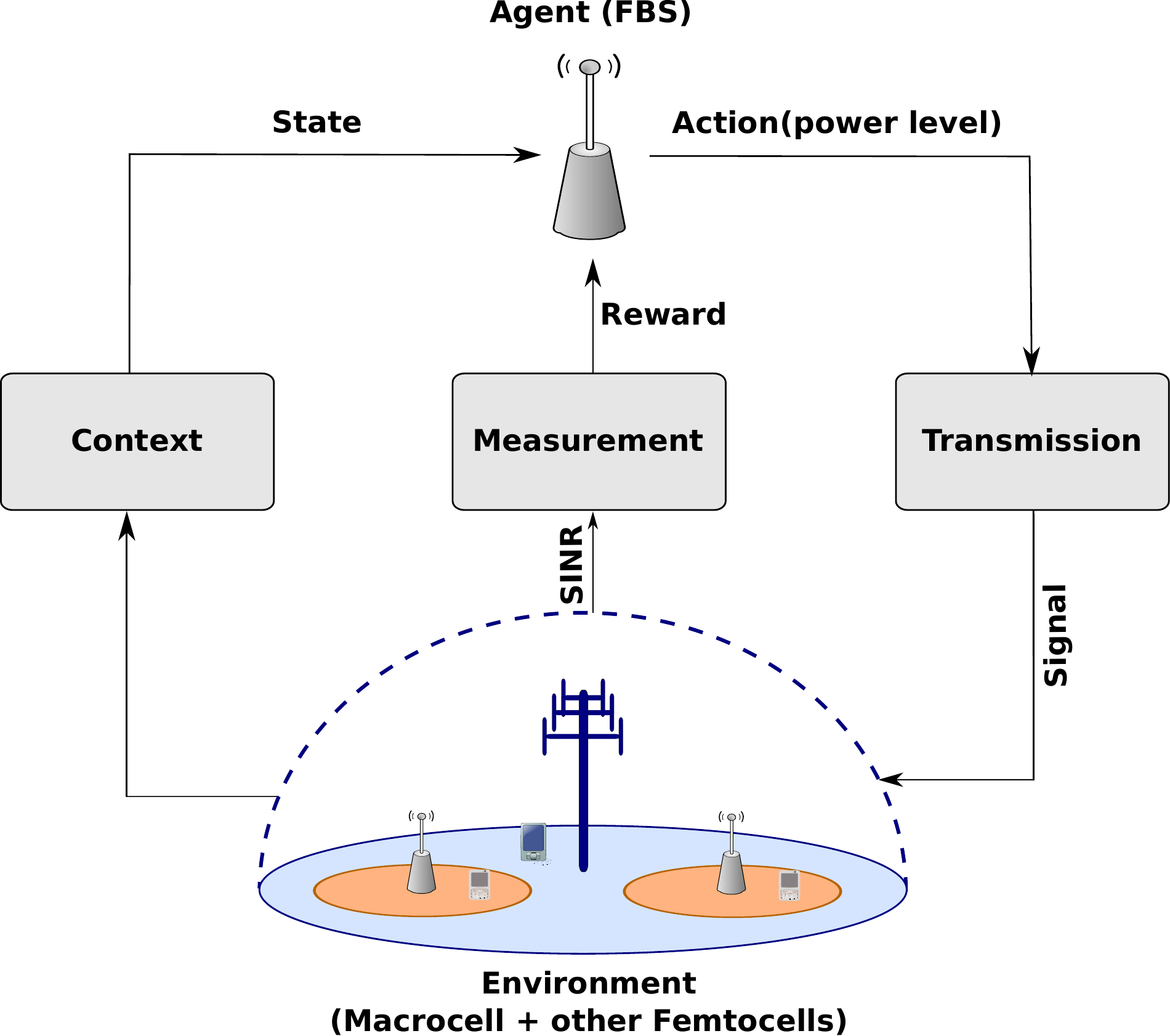}
\caption[width=.3\textwidth]{\small The proposed learning framework: the environment from the point of view of an agent (FBS), and its interaction with the environment in the learning procedure. Context defines the data needed to derive the state of the agent. Measurement refers to calculations needed to derive the reward of the agent.}
\label{fig_framework}
\end{centering}
\end{figure}

In a wireless communication system, the resource management policy is equivalent to the policy function in an MDP. To integrate the femtocell network in a multi-agent MDP, we define the followings according to Fig.~\ref{fig_framework}.

\begin{itemize}[leftmargin=*]
\item \textbf{Environment}: From the view point of an FBS, the environment is comprised of the macrocell and all other femtocells.

\item \textbf{Agent}: Each FBS is an independent agent in the MDP. In this paper, the terms of agent and FBS are used interchangeably. An agent has three objectives: (\rmnum{1}) improving its sum transmission rate, (\rmnum{2}) guaranteeing the required SINR for its user (i.e., $\Gamma_k$), and (\rmnum{3}) meeting the required SINR for the MUE.

\item \textbf{Action set} ($\mathcal{A}_k$): The transmit power level is the action of an FBS. The $k$th FBS chooses its transmit power from the set $\mathcal{A}_k$ which covers the space between $\textit{p}_\text{min}$ and $\textit{p}_\text{max}$. $\textit{p}_\text{min}$ and $\textit{p}_\text{max}$ denote the minimum and maximum transmit power of the FBS, respectively. In general, the FBS has no knowledge of the environment and it chooses its actions with the same probability in the training mode. Therefore, equal step sizes of $\Delta p$ are chosen between $p_{min}$ and $p_{max}$ to construct the set $\mathcal{A}_k$.

\item \textbf{State set} ($\mathcal{X}_k$): State set directly affects the performance of the MUE and the FUEs. To this end, we define four variables to represent the state of the network. The state set variables are defined based on the constraints of the optimization problem in~\eqref{eq_opt}. We define the variables $X_1$ and $X_2$ as indicators of the performance of the FUE and the MUE. On the other hand, the relative location of an FBS with respect to the MUE and the MBS is important and affects the interference power at the MUE caused by the FBS, and the interference power at the FBS causes by the MBS. Therefore, we define $X_3$ as an indicator of the interference imposed on the MUE by the FBS, and $X_4$ as an indicator of interference imposed on the femtocell by the MBS. The state variables are defined as
\begin{itemize}
\item {$X_1 \in \left\lbrace 0, 1\right\rbrace$}: The value of $X_1$ indicates whether the FBS is supporting its FUE with the required minimum SINR or not. $X_1$ is defined as $X_1=\mathbbm{1}_{\left\lbrace \gamma_k \geq \Gamma_k \right\rbrace}$.
\item {$X_2 \in \left\lbrace 0, 1\right\rbrace$}: The value of $X_2$ indicates whether the MUE is being supported with its required minimum SINR or not. $X_2$ is defined as $X_2=\mathbbm{1}_{\left\lbrace \gamma_0 \geq \Gamma_0\right\rbrace}$.
\item {$X_3 \in \left\lbrace 0,1,2,...,N_1 \right\rbrace$}: The value of $X_3$ defines the location of the FBS compared to $N_1$ concentric rings around the MUE. The radius of rings are $d_{1}$, $d_2$, ... , $d_{N_1}$.
\item {$X_4 \in \left\lbrace 0,1,2,...,N_2 \right\rbrace$}: The value of $X_4$ defines the location of the FBS compared to $N_2$ concentric rings around the MBS. The radius of rings are $d'_1$, $d'_2$, ... , $d'_{N_2}$.
\end{itemize}
The $k$th FBS calculates $\gamma_k$ based on the channel equality indicator (CQI) received from its related FUE to assess $X_1$. The MBS is aware of the SINR of the MUE user, i.e., $\gamma_0$, and the relative location of the FBS concerning itself and the MUE. Therefore, the FBS obtains the required information to asses the $X_2$, $X_3$, and $X_4$ variables via backhaul and feedback from the MBS.

Here, we defined the state variables as a function of each FBS's SINR and location. Therefore, in high SINR regime, the state of FBSs can be assumed to be independent of each other.

In Section~\ref{sec_sim}, we will examine different possible state sets to investigate the effect of the above state variables on the performance of the network.
\end{itemize}
\section{Q-DPA, Reward Function, and Sample Complexity} \label{sec_QDPA}
In this section, we present Q-DPA, which is an application of the proposed framework. Q-DPA details the learning method, the learning rate, and the training procedure. Then, the proposed reward function is defined. Finally, the required sample complexity for the training is derived.
\vspace{-10pt}
\subsection{Q-learning Based Distributed Power Allocation (Q-DPA)}\label{sec_QDPA}

To solve the Bellman equation in~\eqref{eq_Bellman}, we use Q-\textit{learning}. The reasoning for choosing the RL method and advantages of Q-\textit{learning} are explained in Sections~\ref{sec_MDP} and~\ref{sec_litSurvey}, respectively. The Q-\textit{learning} update rule to evaluate a policy for the global Q-function can be represented as~\cite{Watkins1992}
\begin{align}\label{eq-qlearning}
\mathbf{Q}(\mathbf{x}^{\left(t\right)}, \mathbf{a}^{\left(t\right)}) \leftarrow \mathbf{Q}(\mathbf{x}^{\left(t\right)}, \mathbf{a}^{\left(t\right)}) + \alpha^{\left(t\right)}\left(\mathbf{x}, \mathbf{a} \right) \left(\mathbf{R}^{\left(t+1\right)}\left(\mathbf{x}^{\left(t\right)}, \mathbf{a}^{\left(t\right)} \right) + \beta\underbrace{\underset{\mathbf{a^\prime}}\max~\mathbf{Q}(\mathbf{x}^{\left(t+1\right)}, \mathbf{a'})}_{\left(M\right)}-\mathbf{Q}(\mathbf{x}^{\left(t\right)}, \mathbf{a}^{\left(t\right)})\right),
\end{align}
where $\mathbf{a^\prime} \in \mathcal{A}$, $\alpha^{\left(t\right)}\left(\mathbf{x}, \mathbf{a} \right)$ denotes the learning rate 
at time step $t$, and $\mathbf{x}^{\left(t\right)}$ is the new state of the network. The term $M$ is the maximum value of the global Q-function that is available at the new state $\mathbf{x}^{\left(t+1\right)}$. After each iteration, the FBSs will receive the delayed reward $\mathbf{R}^{\left(t+1\right)}\left(\mathbf{x}^{\left(t\right)}, \mathbf{a}^{\left(t\right)} \right)$ and then the global Q-function will be updated according to~\eqref{eq-qlearning}. 


In the prior works~\cite{art_reward1, art_reward2, art_reward, art_femto, art_QL_ISJ}, a constant learning rate was used for Q-\textit{learning} to solve the required optimization problems. However, according to~\cite{art_learning_rate}, in finite number of iterations, the performance of Q-\textit{learning} can be improved by applying a decaying learning rate. 
Therefore, we use the following learning rate
\vspace{-10pt}
\begin{align} \label{eq_learning_rate}
\alpha^{\left(t\right)}\left(\mathbf{x}, \mathbf{a} \right) = \frac{1}{\left[ 1+t\left(\mathbf{x}, \mathbf{a} \right)\right]},
\end{align}
in which $t\left(\mathbf{x}, \mathbf{a} \right)$ refers to the number of times, until time step $t$, that the state-action pair $\left(\mathbf{x}, \mathbf{a} \right)$ is visited. It is worth mentioning that, by using the above learning rate, we need to keep track of the number of times each state-action pair has been visited during training, which requires more memory. Therefore, at the cost of more memory, a better performance can be achieved.

There are two alternatives available for the training of new FBSs as they join the network, they can use independent learning or cooperative learning. In independent learning, each FBS tries to maximize its own Q-function. In other words, using the factorization method in Section~\ref{sec_factored}, the term $M$ in~\eqref{eq-qlearning} is approximated as
\vspace{-8pt}
\begin{align}\label{eq_IL}
M = \underset{\mathbf{a}^\prime}\max \sum\limits_{k \in \mathcal{K}} Q_k(\mathbf{x}_k^{\left(t+1\right)},a_k^\prime) \approx \sum\limits_{k \in \mathcal{K}} \underset{a_k^\prime}\max~Q_k\left(\mathbf{x}_k^{\left(t+1\right)},a_k^\prime\right).
\end{align}
In cooperative learning, the FBSs share their local Q-functions and will assume that the FBSs with the same state make the same decision. 
Hence, term $M$ is approximated as
\vspace{-8pt}
\begin{align}\label{eq_CL_rule0}
M = \underset{\mathbf{a}^\prime}\max \sum\limits_{k \in \mathcal{K}} Q_k(\mathbf{x}_k^{\left(t+1\right)},a^\prime_k) \approx \underset{a_k^\prime}\max \sum\limits_{k \in \mathcal{K}^\prime} Q_k\left(\mathbf{x}_k^{\left(t+1\right)}, a_k^\prime\right),
\end{align}
where $\mathcal{K}^\prime$ is the set of FBSs with the same state $\mathbf{x}_k^{\left(t+1\right)}$. Cooperative Q-\textit{learning} may result in a higher cumulative reward~\cite{art_marl_survey}. 
 However, cooperation will result in the same policy for FBSs with the same state and additional overhead since the Q-functions between FBSs need to be shared over the backhaul network. The local update rule for the $k$th FBS can be derived from~\eqref{eq-qlearning} as
\begin{align}\label{eq_local}
Q_k(\mathbf{x}_k^{\left(t\right)}, a_{k}^{\left(t\right)}) \leftarrow Q_k(\mathbf{x}_k^{\left(t\right)}, a_{k}^{\left(t\right)}) + \alpha^{\left(t\right)} \left(R^{\left(t+1\right)}\left(\mathbf{x}_k^{\left(t\right)}, a_k^{\left(t\right)} \right) + \beta Q_k\left(\mathbf{x}_k^{\left(t+1\right)},a_k^*\right) -Q_k(\mathbf{x}_k^{\left(t\right)}, a_{k}^{\left(t\right)})\right),
\end{align}

where, $R^{\left(t+1\right)}\left(\mathbf{x}_k^{\left(t\right)}, a_k^{\left(t\right)} \right)$ is the reward of the $k$th FBS, and $a_k^*$ is defined as
\begin{align}
\underset{a_k^\prime} \argmax~Q_k\left(\mathbf{x}_k^{\left(t+1\right)},a_k^\prime\right),
\end{align}
\vspace{-8pt}
and
\vspace{-8pt}
\begin{align}
\underset{a_k^\prime} \argmax \sum\limits_{k \in \mathcal{K}^\prime} Q_k\left(\mathbf{x}_k^{\left(t+1\right)}, a_k^\prime\right),
\end{align}
for independent and cooperative learning, respectively.

In this paper, a tabular representation is used for the Q-function in which the rows of the table refer to the states and the  columns refer to the actions of an agent. Generally, for large state spaces, neural networks are more efficient to use as Q-functions, however, part of this work is focused on the effect of state space variables. Therefore, we avoid large number of state variables. On the other hand, we provide exhaustive search solution to investigate the optimality of our solution which is not possible for large state spaces.

The training for an FBS happens over $L$ frames. 
In the beginning of each frame, the FBS chooses an action, i.e., transmit power. 
Then, the FBS sends a frame to the intended FUE. The FUE feeds back the required measurements such as CQI so the FBS can estimate the SINR at the FUE, and calculate the reward based on~\eqref{RF2}. Finally, the FBS updates its Q-table according to~\eqref{eq_local}.

Due to limited number of training frames, each FBS needs to select its actions in a way that covers most of the action space and improves the policy at the same time. Therefore, the FBS chooses the actions with a combination of exploration and exploitation, known as an $e$-greedy exploration. In the $e$-greedy method, the FBS acts greedily with probability $1-e$ (i.e., exploiting) and randomly with probability $e$ (i.e., exploring). In exploitation, the FBS selects an action that has the maximum value in the current state in its own Q-table (independent learning) or in the summation of Q-tables (cooperative learning). In exploring, the FBS selects an action randomly to cover action space and avoid biasing to a local maximum. In~\cite{book_sutton}, it is shown that for a limited number of iterations the $e$-greedy policy results in a closer final value to the optimal value compared to only exploiting or exploring. 


It is worth mentioning that the overhead of sharing Q-tables depends on the definition of the state model $\mathcal{X}_k$ according to Section~\ref{sec_femtocell_MDP}. For instance, assuming the largest possible state model as $\mathcal{X}_k=\left\lbrace X_1, X_2, X_3, X_4\right\rbrace$. The variables $X_3$ and $X_4$ depend on the location of the FBS and are fixed during training. Therefore, one training FBS uses four rows of its Q-table and just needs the same rows from other FBSs. Hence, if the number of active FBSs is $|\mathcal{K}|$, the number of messages to the FBS in each training frame is $4\times \left(|\mathcal{K}|-1\right)$, each of size $|\mathcal{A}_k|$.
 
\subsection{Proposed Reward Function}\label{sec_reward_design} 

The design of the reward function is essential because it directly impacts the objectives of the FBS. Generally, there has not existed a quantitative approach to designing the reward function. Here, we present a systematic approach for deriving the reward function based on the nature of the optimization problem under consideration. Then, we compare the behavior of the designed reward function to the ones in~\cite{art_reward1, art_reward2, art_reward}. 

The reward function for the $k$th FBS is represented as $R_k$. According to the Section~\ref{sec_femtocell_MDP}, the $k$th FBS has knowledge of the minimum required SINR for the MUE, i.e. $\Gamma_0$, and minimum required SINR for its related FUE, i.e. $\Gamma_k$. Also, after taking an action in each step, the $k$th FBS has access to the rate of the MUE, i.e. $r_0$ and the rate of its related FUE, i.e. $r_k$. Therefore, $R_k$ is considered as a function of the above four variables as $R_k: \left(r_0, r_k, \Gamma_0, \Gamma_k\right) \rightarrow \mathbb{R}$.
	

In order to design the appropriate reward function, we need to estimate the progress of the $k$th FBS toward the goals of the optimization problem. 
Based on the input arguments to the reward function, we define two progress estimators, one for the MUE as $\left( r_0-\log_2\left(1+\Gamma_0\right) \right)$ and one for the $k$th FUE as $\left( r_k-\log_2\left(1+\Gamma_k\right) \right)$. To reduce computational complexity, we define the reward function as a polynomial function of the defined progress estimators as
\begin{align}\label{eq_rf_0}
\begin{split}
  R_k\left(r_0, r_k, \Gamma_0, \Gamma_k\right) =
     \left( r_0-\log_2\left(1+\Gamma_0\right)\right)^{k_1} + \left( r_k-\log_2\left(1+\Gamma_k\right) \right)^{k_2} + C,
\end{split}
\end{align}
where, $k_1$ and $k_2$ are integers and $C \in \mathbb{R}$ is a constant referred to as the bias of the reward function.

The constant bias, $C$, in the reward function has two effects on the learning algorithm: (\rmnum{1}) The final value of the states for a given policy $\pi$, and (\rmnum{2}) the behavior of the agent in the beginning of the learning process as follows:
\begin{enumerate}[leftmargin=*]
\item Effect of bias on the final value of the states: Assume the reward function, $R_1=f\left(\cdot \right)$, and the reward function $R_2=f\left(\cdot \right) + C$, $C \in \mathbb{R}$. 
 We define the value of state $\mathbf{x}$ for a given policy $\pi$ using $R_1$ as $V_{1}\left(\mathbf{x}\right)$ and the value of the state $\mathbf{x}$ for the same policy using $R_2$ as $V_{2}\left(\mathbf{x}\right)$. According to~\eqref{eq_value_func}
\begin{align}
V_{2}\left(\mathbf{x}\right) = \mathbb{E}_{\pi} \left[ \sum_{t=0}^{\infty} \beta^t \left( f^{\left(t+1\right)}\left(\cdot \right) + C \right)  \right] = \mathbb{E}_{\pi} \left[ \sum_{t=0}^{\infty} \beta^t f^{\left(t+1\right)}\left(\cdot \right) \right] + C \sum_{t=0}^{\infty} \beta^t = V_{1}\left(\mathbf{x}\right) + \frac{C}{1-\beta}.
\end{align}
Therefore, bias of the reward function adds the constant value $\frac{C}{1-\beta}$ to the value of the states. However, all the states are affected the same after the convergence of the algorithm.
\item Effect of bias in the beginning of the learning process: This effect is studied using the action-value function of an agent, i.e., the Q-function. Assume that the Q-function of the agent is initialized with zero values and the reward function is defined as $R=f\left(\cdot \right) + C$. Further let us consider the first transition of the agent from state $\mathbf{x}^\prime$ to state $\mathbf{x}^{\prime\prime}$ happens by taking action $a$ at time step $t$, i.e., $\mathbf{x}^{\left(t\right)} = \mathbf{x}^\prime$ and $\mathbf{x}^{\left(t+1\right)} = \mathbf{x}^{\prime\prime}$. The update rule at time step $t$ is given by~\eqref{eq_local}
\vspace{-10pt}
\begin{align}
	\begin{split}
Q(\mathbf{x}^\prime, a) &\leftarrow Q(\mathbf{x}^\prime, a) + \alpha^{\left(t\right)}\left(\mathbf{x}^\prime, a \right) \left(R\left(\mathbf{x}^\prime, a \right) + \beta~\underset{a^\prime} \max~Q\left(\mathbf{x}^{\prime\prime},a^\prime\right) -Q(\mathbf{x}^\prime, a)\right) \\
&\leftarrow \alpha^{\left(t\right)}\left(\mathbf{x}^\prime, a \right) \left( f\left(\cdot \right) + \beta~\underset{a^\prime} \max~Q\left(\mathbf{x}^{\prime\prime},a^\prime\right)\right) + \underbrace{\alpha^{\left(t\right)}\left(\mathbf{x}^\prime, a \right) C}_{(A)}.
	\end{split}
\end{align}
According to the above, after the first transition from the state $\mathbf{x}^{\prime}$ to the state $\mathbf{x}^{\prime\prime}$, the Q-value for the state $\mathbf{x}^{\prime}$ is biased by the term (A). If ($A >0$), the value of the state $\mathbf{x}^{\prime}$ increases and if ($A <0$), the value of the state $\mathbf{x}^{\prime}$ decreases. Therefore, the already visited states will be more or less attractive to the agent in the beginning of the learning process as long as the agent has not explored the state-space enough. 
\end{enumerate}
The change of behavior of the agent in the learning process can be used to bias the agent towards the desired actions or states. However, in basic Q-learning the agent has no knowledge in prior about the environment. Therefore, we select the bias equal to zero, $C=0$, and define the reward function as
\begin{definition}\label{def_reward}
The reward function for the $k$th FBS, $R_k: \left(r_0, r_k, \Gamma_0, \Gamma_k\right) \rightarrow \mathbb{R}$, is a continuous and differentiable function on $\mathbb{R}^2$ defined as
\vspace{-10pt}
\begin{align}
R_k\left(r_0, r_k, \Gamma_0, \Gamma_k\right) = \left( r_0-\log_2\left(1+\Gamma_0\right)\right)^{k_1} + \left( r_k-\log_2\left(1+\Gamma_k\right) \right)^{k_2},
\end{align}
where $k_1$ and $k_2$ are integers.
\end{definition}



The objective of the FBS is to maximize its transmission rate. On the other hand, high transmission rate for the MUE is a priority for the FBS. Therefore, $R_k$ should have the following property
\vspace{-5pt}
\begin{align}\label{eq_cond3}
\frac{\partial R_k}{\partial r_i} \geq 0,~i=0,k.
\end{align}

The above property implies that higher transmission rate for the FBS or the MUE results in higher reward. Hence, considering Definition~\ref{def_reward}, we design a reward function that motivates the FBSs to increase $r_k$ and $r_0$ as much as possible even more than the required rate as follow
\vspace{-5pt}
\begin{align}\label{RF2}
\begin{split}
  R_k=
     \left( r_0-\log_2\left(1+\Gamma_0\right)\right)^{2m-1} + \left( r_k-\log_2\left(1+\Gamma_k\right) \right)^{2m-1},
\end{split}
\end{align}
where $m$ is an integer. The above reward function considers the minimum rate requirements of the FUE and the MUE, while encourages the FBS to increase transmission rate of both.

To further understand the proposed reward function, we discuss reward functions that are used by~\cite{art_reward1, art_reward2, art_reward}. We refer to the designed reward function in~\cite{art_reward1} as quadratic, in~\cite{art_reward2} as exponential, and in~\cite{art_reward} as proximity reward functions. The quadratic reward function is designed based on a conservative approach. In fact, the FBS is enforced to select actions that result in transmission rate close to the minimum requirement. Therefore, higher or lower rate than the minimum requirement results in a same amount of reward. The behavior of the quadratic reward function can be explained as follow
\begin{align}\label{eq_cond1}
\frac{\partial R_k}{\partial r_i} \times \left( r_i - \log_2\left(1+\Gamma_i\right)\right) \leq 0,~i=0,k.
\end{align}
The above property implies that if the rate of the FBS or the MUE is higher than the minimum requirement, the actions that increase the rate will decrease the reward. Hence, this property is against increasing sum transmission rate of the network. The exponential and proximity reward functions have the property in~\eqref{eq_cond3} for the rate of the FBS, and the property in~\eqref{eq_cond1} for the rate of the MUE. In another words, they satisfy the following properties
\vspace{-10pt}
\begin{equation}\label{eq_cond2}
\begin{aligned}
&\frac{\partial R_k}{\partial r_0} \times \left( r_0 - \log_2\left(1+\Gamma_0\right)\right) \leq 0,\\
&\frac{\partial R_k}{\partial r_k} \geq 0.
\end{aligned}
\end{equation}
As the density of the FBSs increases, the above properties result in increasing transmit power to achieve higher individual rate for a FUE while introducing higher interference for the MUE and other neighbor FUEs. In fact, as increasing the FUE rate is rewarded, taking actions that result in increasing the MUE rate decreases the reward. However, the FBS should have the option of decreasing its transmit power to increase the rate of the MUE. This behavior is important since it causes an FBS to produce less interference for its neighboring femtocells. Therefore, we give equal opportunity for increasing the rate of the MUE or the FUE.

\begin{figure}
    \centering
    \begin{subfigure}[t]{.350\textwidth}
        \centering
        \includegraphics[width=1.0\columnwidth]{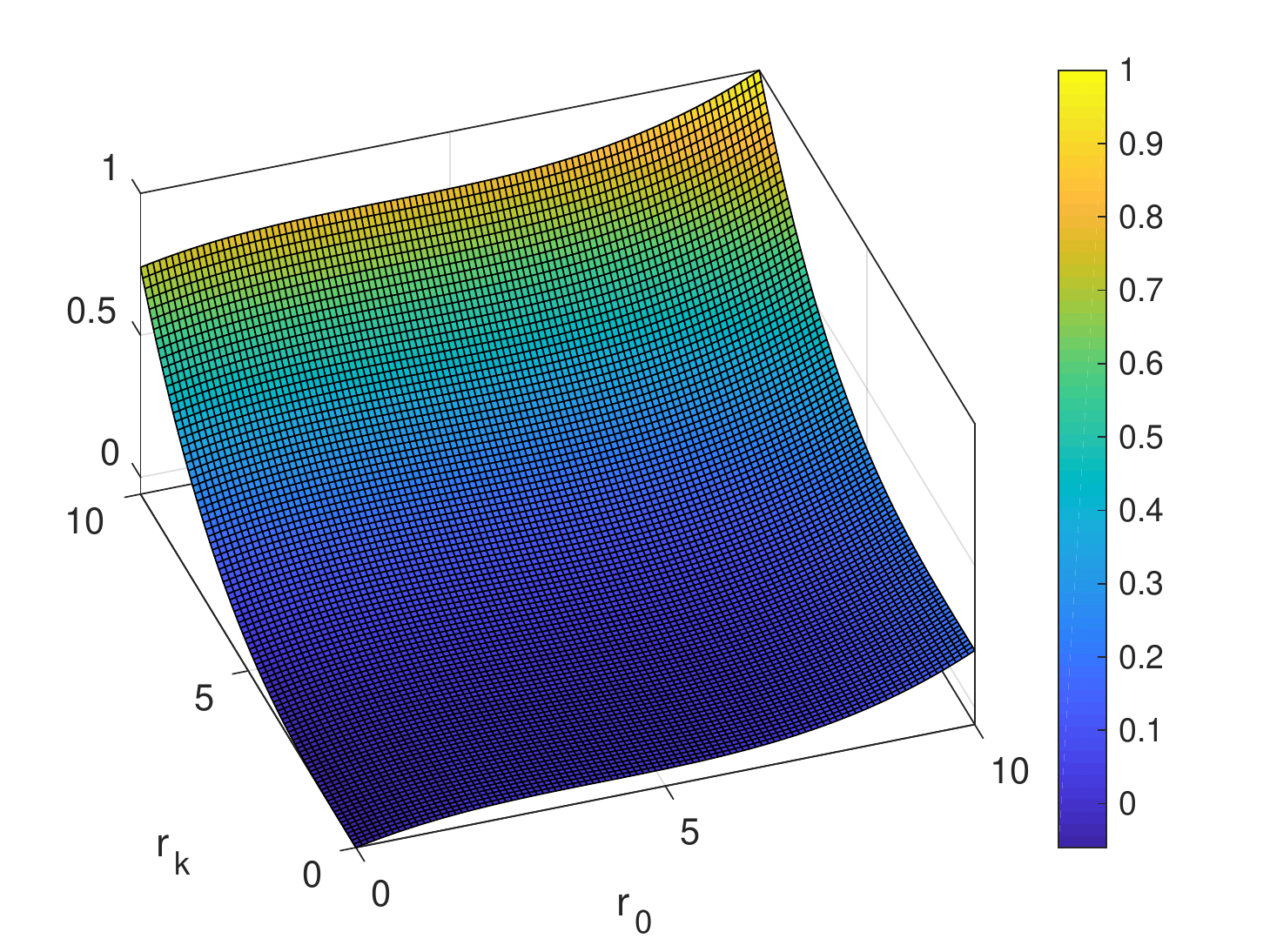}
        \caption{}\label{fig:RF_prop}
    \end{subfigure}%
    \begin{subfigure}[t]{.350\textwidth}
        \centering
        \includegraphics[width=1.0\columnwidth]{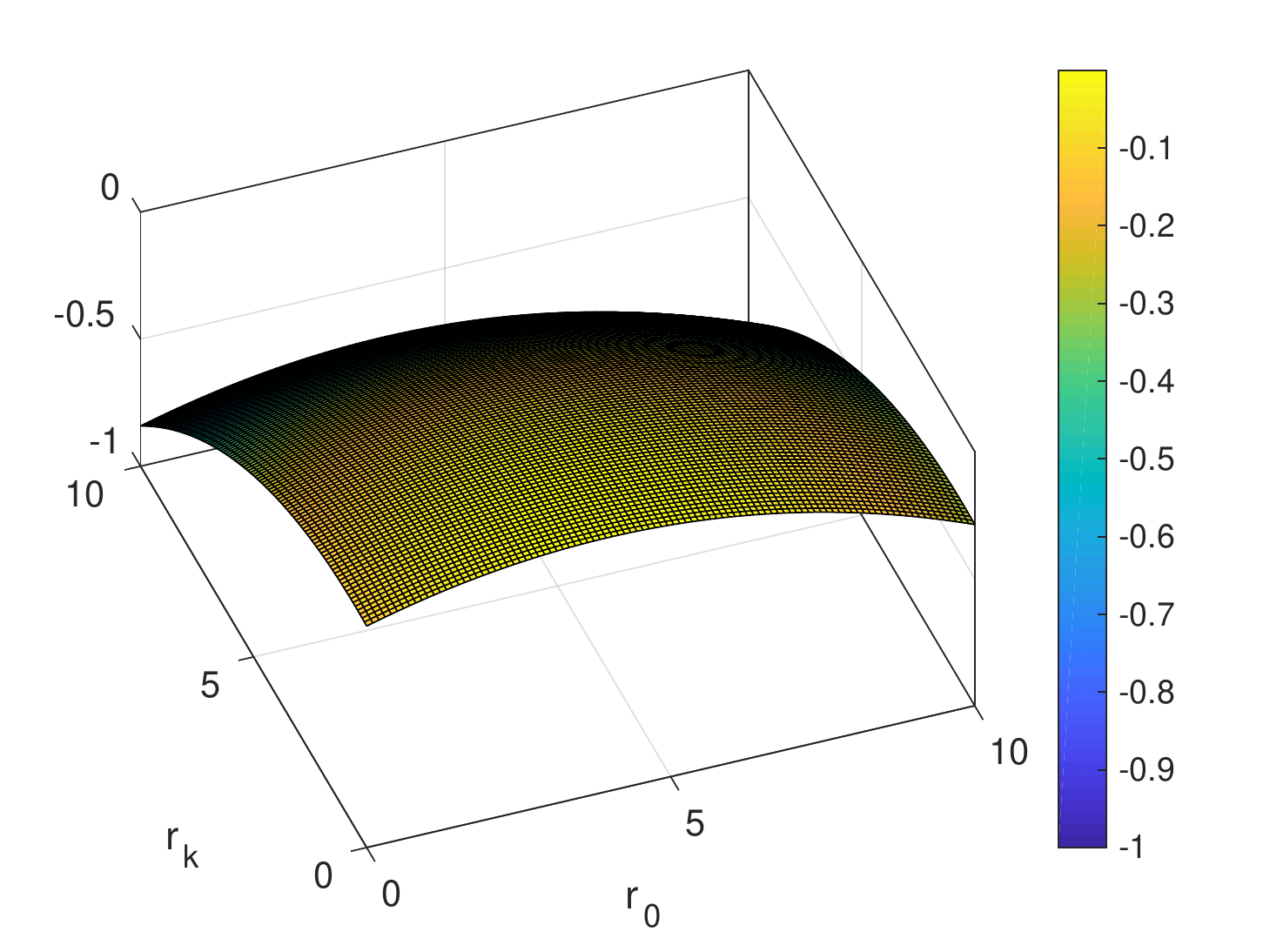}
        \caption{}\label{fig:RF_quad}
    \end{subfigure}
    \begin{subfigure}[t]{.350\textwidth}
        \centering
        \includegraphics[width=1.0\columnwidth]{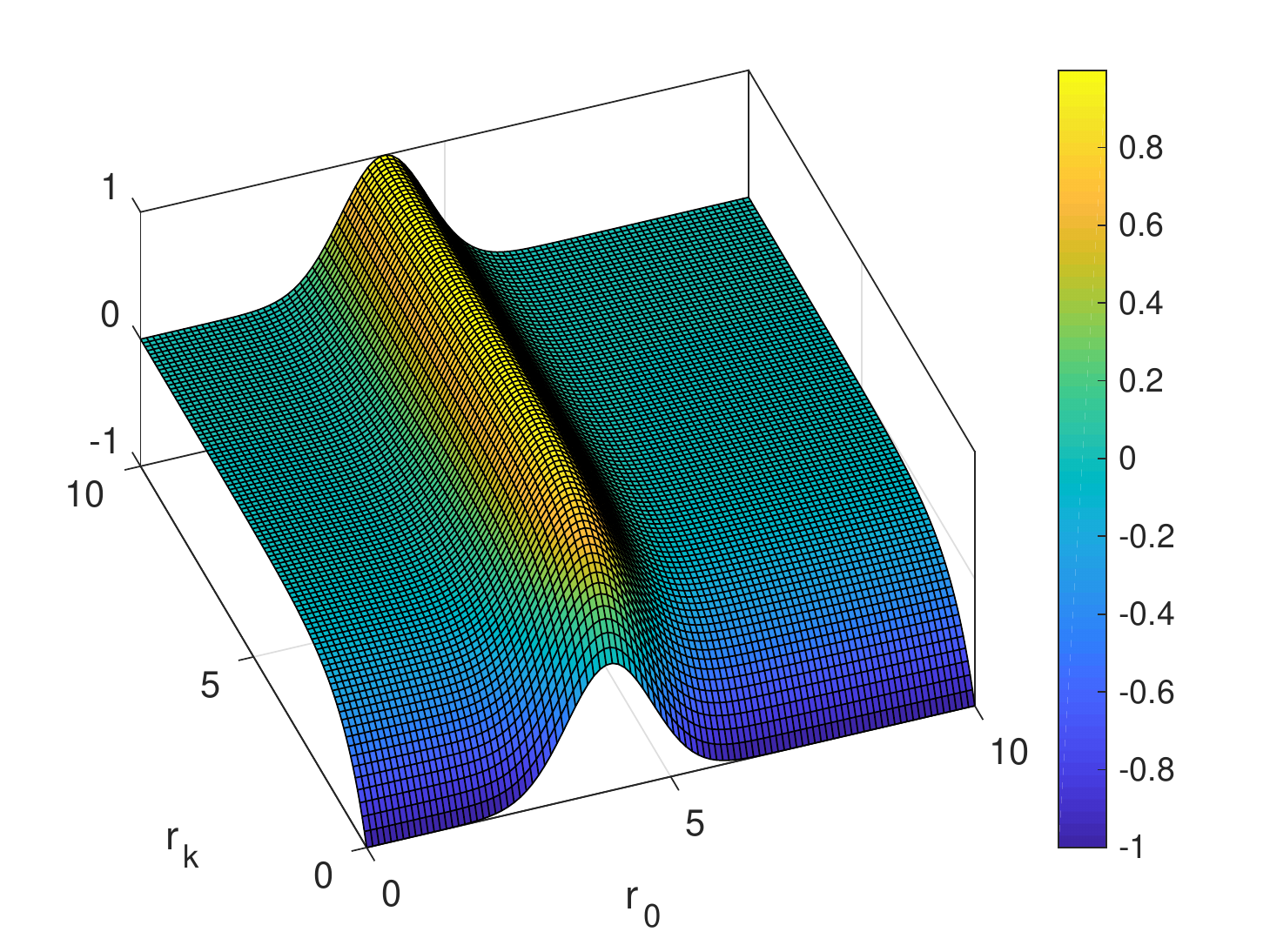}
        \caption{}\label{fig:RF_exp}
    \end{subfigure}%
    \begin{subfigure}[t]{.350\textwidth}
        \centering
        \includegraphics[width=1.0\columnwidth]{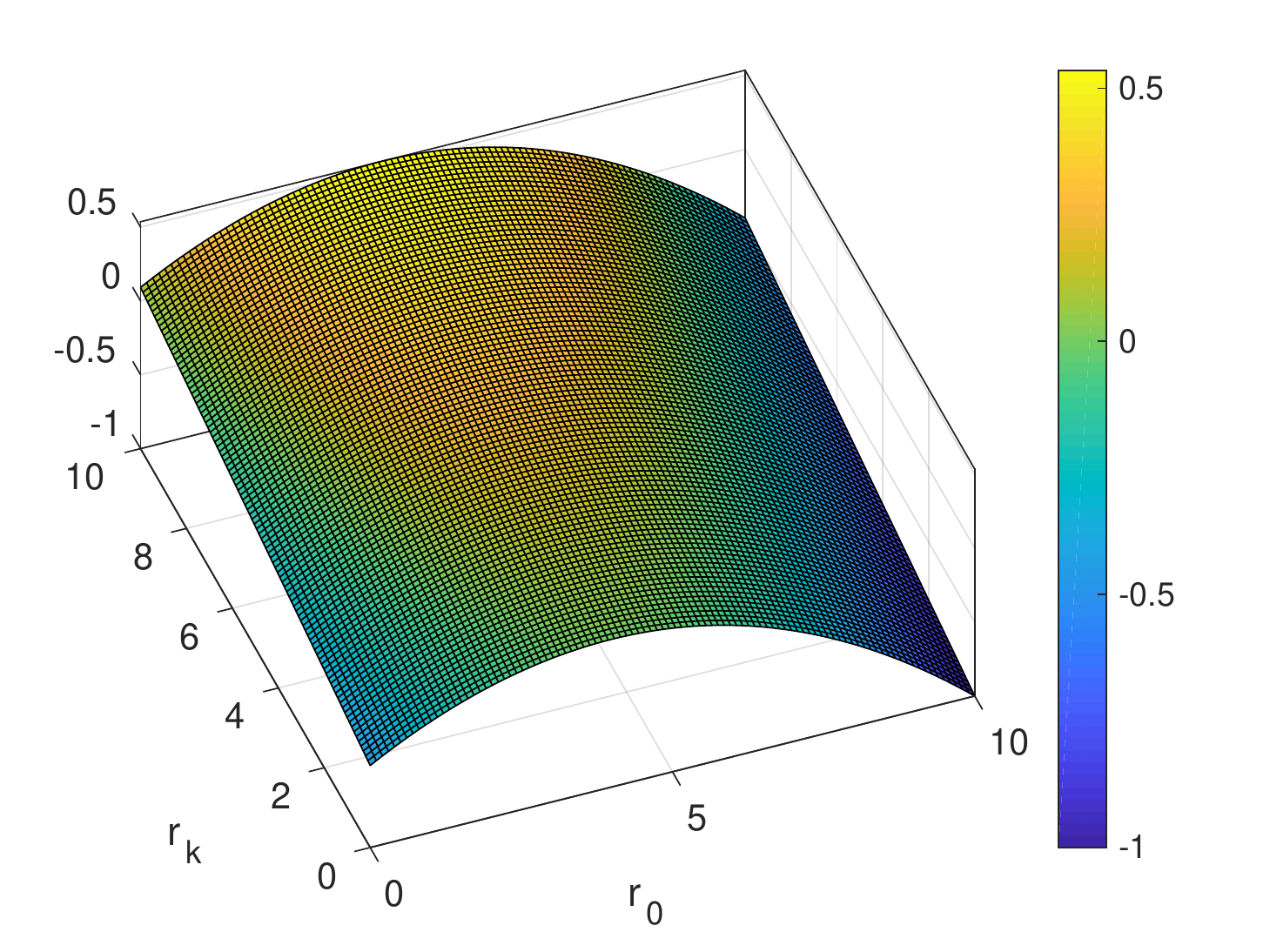}
        \caption{}\label{fig:RF_prox}
    \end{subfigure}
        \caption{Reward functions: (a) Proposed reward function with $m=2$, (b) Quadratic reward function with zero maximum at ($4.0, 0.5$), (c) Exponential reward function, (d) Proximity reward function.}
\end{figure}
The value of reward functions for different FBSs is different, however they have the same behavior. Here, we plot the value of the four reward functions that are discussed above. The plots refers to the proposed (Fig.~\ref{fig:RF_prop}), quadratic (Fig.~\ref{fig:RF_quad}), exponential (Fig.~\ref{fig:RF_exp}), and proximity (Fig.~\ref{fig:RF_prox}) reward functions. The important information that can be obtained from these plots are the maximal points of the reward functions, behavior of the reward functions around minimum requirements, and behavior of the reward functions by increasing $r_k$ or $r_0$. The proposed reward function in Fig.~\ref{fig:RF_prop} shows pushing the FBS to select transmit power levels that increase both $r_k$ and $r_0$, while other reward functions have their maximum around the minimum rate requirements.

\subsection{Sample Complexity}\label{sec_sample}
In each training frame, Q-DPA collects one sample from the environment represented as the state-action pair in the Q-function. Sample complexity is defined as the minimum number of samples that is required to train the Q-function to achieve an $\epsilon$-optimal policy. For $\epsilon > 0$ and $\delta \in \left(0,1\right]$, $\pi$ is an $\epsilon$-optimal policy if~\cite{art_convergence1}
\begin{align}\label{eq_SC0}
\Pr\left( \norm{ Q^*-Q_{\pi} }  < \epsilon\right) \geq 1-\delta.
\end{align}
The sample complexity depends on the exploration policy that is generating the samples. In Q-DPA, $e$-greedy policy is used as the exploration policy. However, $e$-greedy policy depends on the Q-function of the agent which is being updated. In fact, the distribution of $e$-greedy policy is unknown. Here, we provide a general bound on the sample complexity of Q-\textit{learning}.

\begin{proposition}\label{theorem1}
Assume $R_{max}$ is the maximum of the reward function for an agent and $Q^{\left(T\right)}$ is the action-value for state-action pair $\left(x,a\right)$ after $T$ iterations. Then, with probability at least $1-\delta$, we have
\begin{align}
\norm{Q^*-Q^{\left(T\right)}} \leq \frac{2 R_{max}}{\left(1-\beta\right)} \left[ \frac{\beta}{T\left(1-\beta\right)}+ \sqrt{\frac{2}{T}\ln{\frac{2\abs{\mathcal{X}}.\abs{\mathcal{A}}}{\delta}}} \right].
\end{align}
\begin{proof}
See Appendix~\ref{appendixTheorem1}.
\end{proof}
\end{proposition}
This proposition proves the stability of Q-\textit{learning} and helps us to provide a minimum number of iterations to achieve $\epsilon > 0$ error with respect to $Q^*$ with probability $1-\delta$ for each state-action pair. By assuming the right term of the above inequality as $\epsilon$, the following Corollary is concluded.

\begin{corollary}\label{corollary1}
For any $\epsilon > 0$, after 
\begin{align}\label{eq_T}
T=\Omega\left(\frac{8R^2_{max}}{\epsilon^2\left(1-\beta\right)^2} \ln{\frac{2\abs{\mathcal{X}} . \abs{A_k}}{\delta}}\right)
\end{align}
number of iterations, $Q^{\left(T\right)}$ reaches $\epsilon$-optimality with probability at least $1-\delta$.
\end{corollary}

\section{Simulation Results} \label{sec_sim}

The objective of this section is to validate the performance of the Q-DPA algorithm with different learning configurations in a dense urban scenario. We first introduce the simulation setup and parameters. Then, we introduce four different learning configurations and we analyze the trade-offs between them. Finally, we investigate the performance of the Q-DPA with different reward functions introduced in Section~\ref{sec_reward_design}. For the sake of simplicity, we use the notation IL as independent learning and CL as cooperative learning.

\subsection{Simulation Setup}\label{sec_setup}
\begin{figure}
\begin{centering}
\includegraphics[width=0.4\columnwidth]{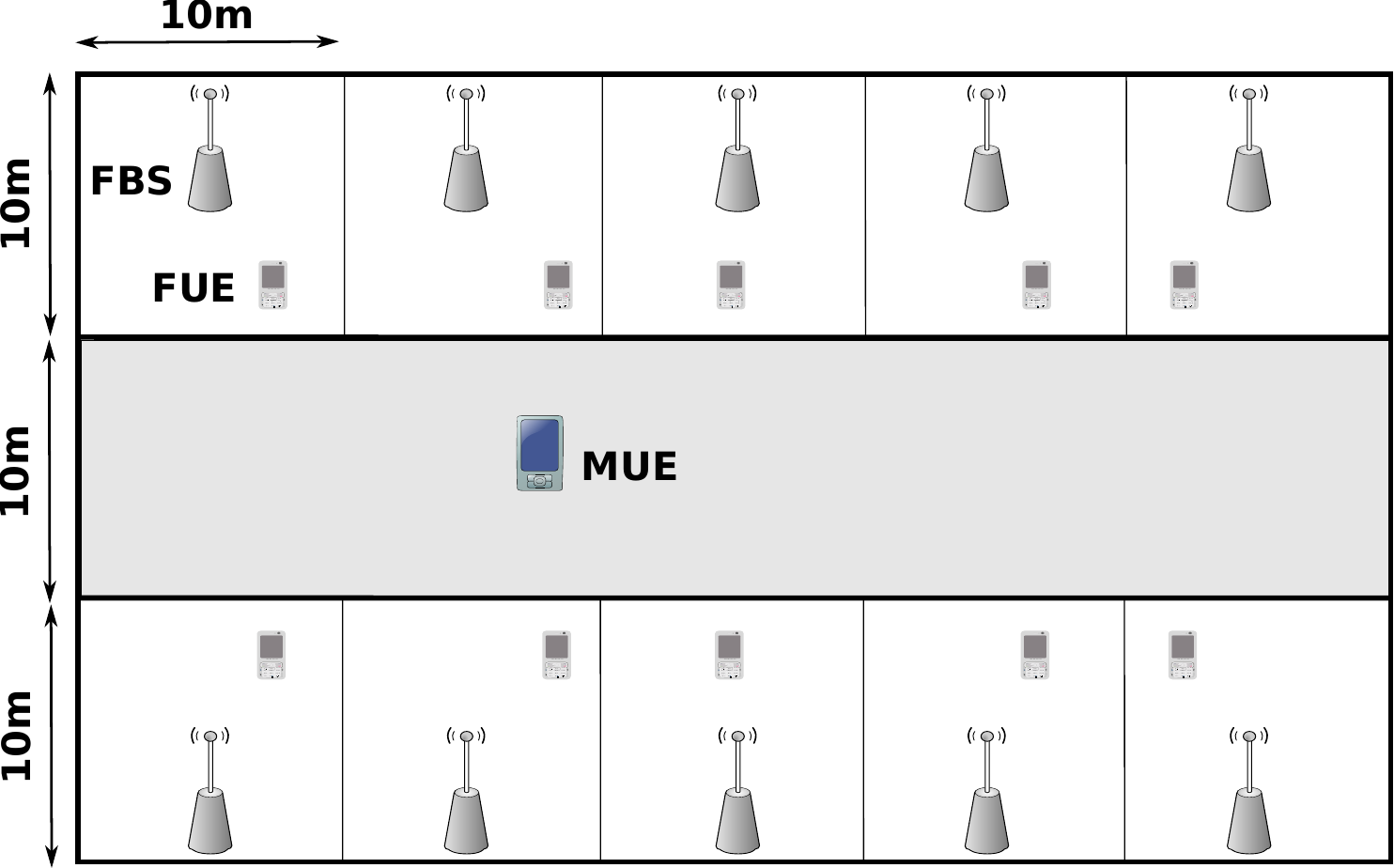}
\caption[width=.3\textwidth]{ Dense urban scenario with a dual strip apartment block located at distance of $350$~m of the MBS; FUEs are randomly located inside each apartment.}
\label{fig_sim}
\end{centering}
\end{figure}

We use a dense urban scenario as the setup of the simulation as illustrated in Fig.~\ref{fig_sim}. We consider one macrocell with radius $350$~m which supports multiple MUEs. The MBS assigns a subband to each MUE. Each MUE is located within a block of apartments and each block contains two strip of apartments. Each strip has five apartments of size $10$~m$\times 10$~m. There is one FBS located in the middle of each apartment which supports an FUE within a $5$~m distance. We assume that the FUEs are always inside the apartments. The FBSs are closed-access, therefore, the MUE is not able to connect to any FBS, however, it receives interference from the FBSs working on the same subband as itself. Here, we assume that the MUE and all the FBSs work on the same sub-carriers to consider the worst case scenario (high interference scenario). However, the extension of the simulation to the multi-carrier scenario is straight forward but does not affect our investigations. We assume the block of apartments is located on the edge of the macrocell, i.e., $350$~m distance from the MBS, and the MUE is assumed to be in between the two strip of apartments.

 In these simulations, in order to initiate the state variables $X_3$ and $X_4$ in Section~\ref{sec_femtocell_MDP}, the number of rings around the MBS and the MUE are assumed to be three ($N_1=N_2=3$). Although, as the density increases, more rings with smaller diameters can be used to more clearly distinguish between the FBSs.

It is assumed that the FBSs and the MBS operate at $f=2.0$ GHz. The MBS allocates $33$~dBm as its transmit power, and the FBSs choose their transmit power from a range of $5$~dBm to $15$~dBm with power steps of $1$~dB. In order to model the pathloss, we use the urban dual strip model from 3GPP TR 36.814~\cite{3gpp.36.814}. The pathloss model of different links are provided in Table~\ref{table_PL}. In Table~\ref{table_PL}, $R$ is the distance between a transmitter and a receiver in meters, $L_{ow}$ is the wall penetration loss which is set to $20$~dB~\cite{3gpp.36.814}. $d_{2D,indoor}$ is the 2-dimensional distance. We assume that the apartments are single floor, therefore, $d_{2D,indoor} \approx R$. The fourth row of the pathloss models is used for the links between the FBSs and the MUE.

\begin{table}
\centering
\setlength\doublerulesep{1.0pt}
\caption{Urban dual strip pathloss model}
\label{table_PL}
\begin{tabular}{llll} \hhline{====} 
   \textbf{Link}                                  & \textbf{PL(dB)} \\ \midrule
	MBS to MUE                                    & \small{$15.3+37.6 \log_{10}R$} ,           \\ 
	MBS to FUE                                    & \small{$15.3+37.6 \log_{10}R+L_{ow}$} ,    \\
	FBS to FUE (same apt strip)                   & \small{$56.76+20 \log_{10}R+0.7d_{2D,indoor}$} , \\ 
	FBS to FUE (different apt strip)              & \small{$max(15.3+37.6 \log_{10}R, 38.46+20 \log_{10}R) + 18.3+0.7d_{2D,indoor}+ L_{ow}$}. \\ \bottomrule
\end{tabular}
\end{table}
The minimum SINR requirements for the MUE and the FUEs are defined based on the required rate needed to support their corresponding user. In our simulations, the minimum required transmission rate to meet the QoS of the MUE is assumed to be $4$ (b/s/Hz), i.e., $\log_2(1+\Gamma_0)=4$ (b/s/Hz). Moreover, for the FUEs the minimum required rate is set to $0.5$ (b/s/Hz), i.e, $\log_2(1+\Gamma_k)=0.5$ (b/s/Hz), $k \in \mathcal{K}$. It is worth mentioning that by knowing the media access control (MAC) layer parameters, the values of the required rates can be calculated using ~\cite[Eqs. (20) and (21)]{art_qos}.

To perform Q-\textit{learning}, the minimum number of required frames, i.e., $L$, is calculated based on achieving $90\%$ optimality, 
with probability of at least $0.9$, i.e., $\delta =0.1$. 
 The simulation parameters are given in Table~\ref{table_1}. The value of the Q-\textit{learning} parameters are selected according to our simulations and references~\cite{art_reward1, art_reward2, art_reward, art_bennis, art_femto, art_QL_ISJ}.
\begin{table}
\centering
\setlength\doublerulesep{1.0pt}
\caption{Simulation Parameters}
\label{table_1}
\begin{tabular}{llll} \hhline{====}
   \textbf{Default parameters} & \textbf{Value} & \textbf{State parameters} & \textbf{Value} \\ \midrule
	Frame time                 & 2 ms           & $d'_1,d'_2,d'_3$          & 50, 150, 400 m \\ 
	UE thermal noise           & -174 dBm/Hz    & $d_1,d_2,d_3$             & 17.5, 22.5, 45 m  \\
	Traffic model              & Fullbuffer     &            \\ \hhline{====}
    \textbf{FBS parameters}    & \textbf{Value} & \textbf{Q-DPA parameters} & \textbf{Value}\\ \midrule
    $\textit{p}_\text{min}$    & 5 dBm          & Training period (iterations) $L$       &$T\times \abs{\mathcal{X}} . \abs{\mathcal{A}_k}$ frames\\
    $\textit{p}_\text{max}$    & 15 dBm         & Learning parameter $\beta$ & 0.9\\
    $\Delta p$                 & 1 dBm          & Exploratory probability ($e$)      & 10\%\\ \bottomrule
    
\end{tabular}
\end{table}

The simulation starts with one femtocell. The FBS starts running Q-DPA in Section~\ref{sec_QDPA} using IL. After convergence, the next FBS is added to the network. The new FBS runs Q-DPA, while the other FBS is already trained, and will just act greedy to choose its transmit power. 
After convergence of the second FBS, the next one is added to the network, and so on. We represent all the results versus the number of active femtocells in the system, from one to ten. Considering the size of the apartment block, and the assumption that all femtocells operate on the same frequency range, the density of deployment varies approximately from $600~\text{FBS}/km^2$ to $6000~\text{FBS}/km^2$.

\subsection{Performance of Q-DPA}\label{sec_qdpa}

\begin{figure}\label{fig:final}
\vspace*{-0.4cm}
    \centering
    \begin{subfigure}[t]{.450\textwidth}
        \centering
        \includegraphics[width=1.0\columnwidth]{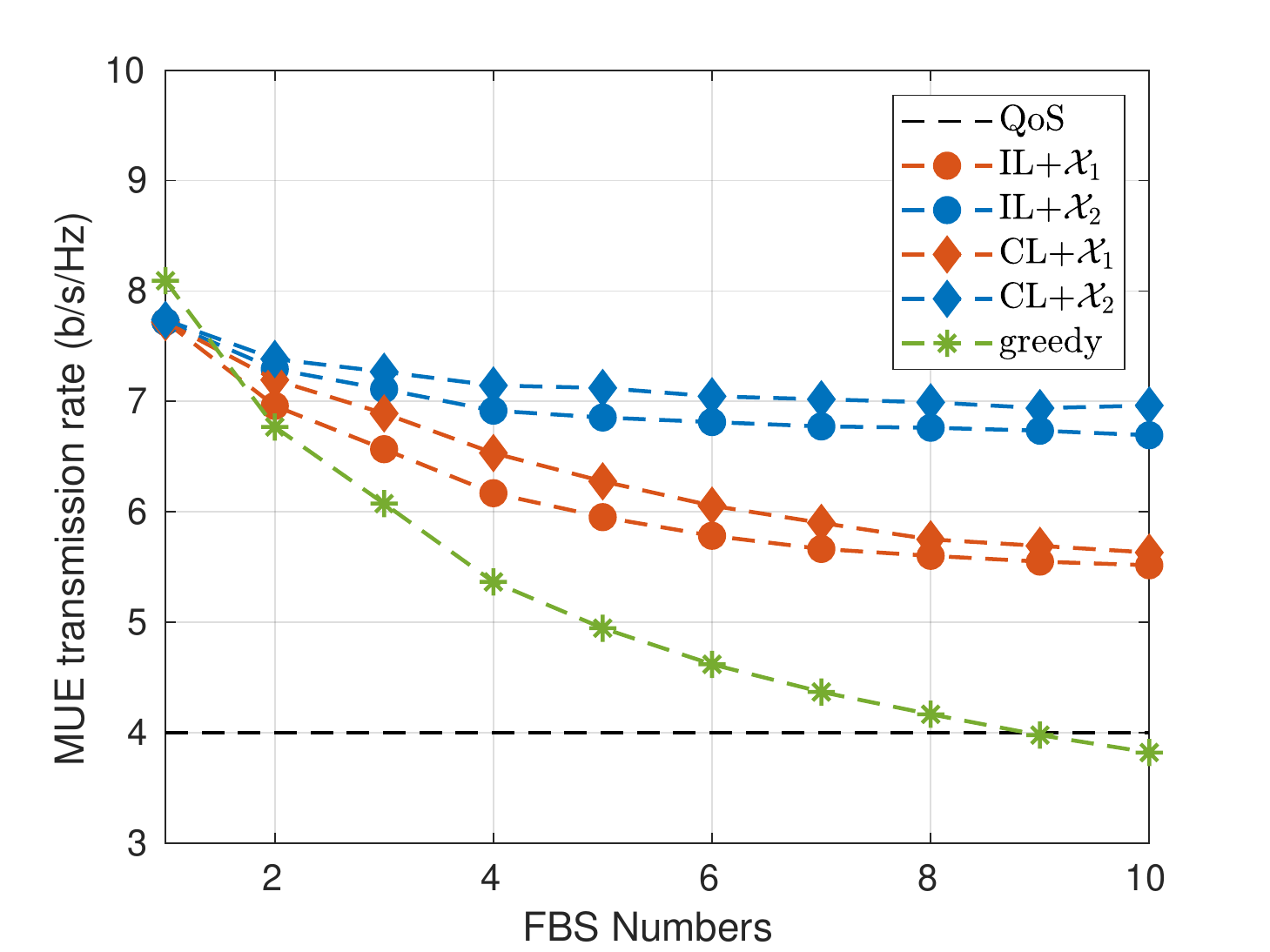}
        \caption{}\label{fig:MUE}
    \end{subfigure}%
    \begin{subfigure}[t]{.450\textwidth}
        \centering
        \includegraphics[width=1.0\columnwidth]{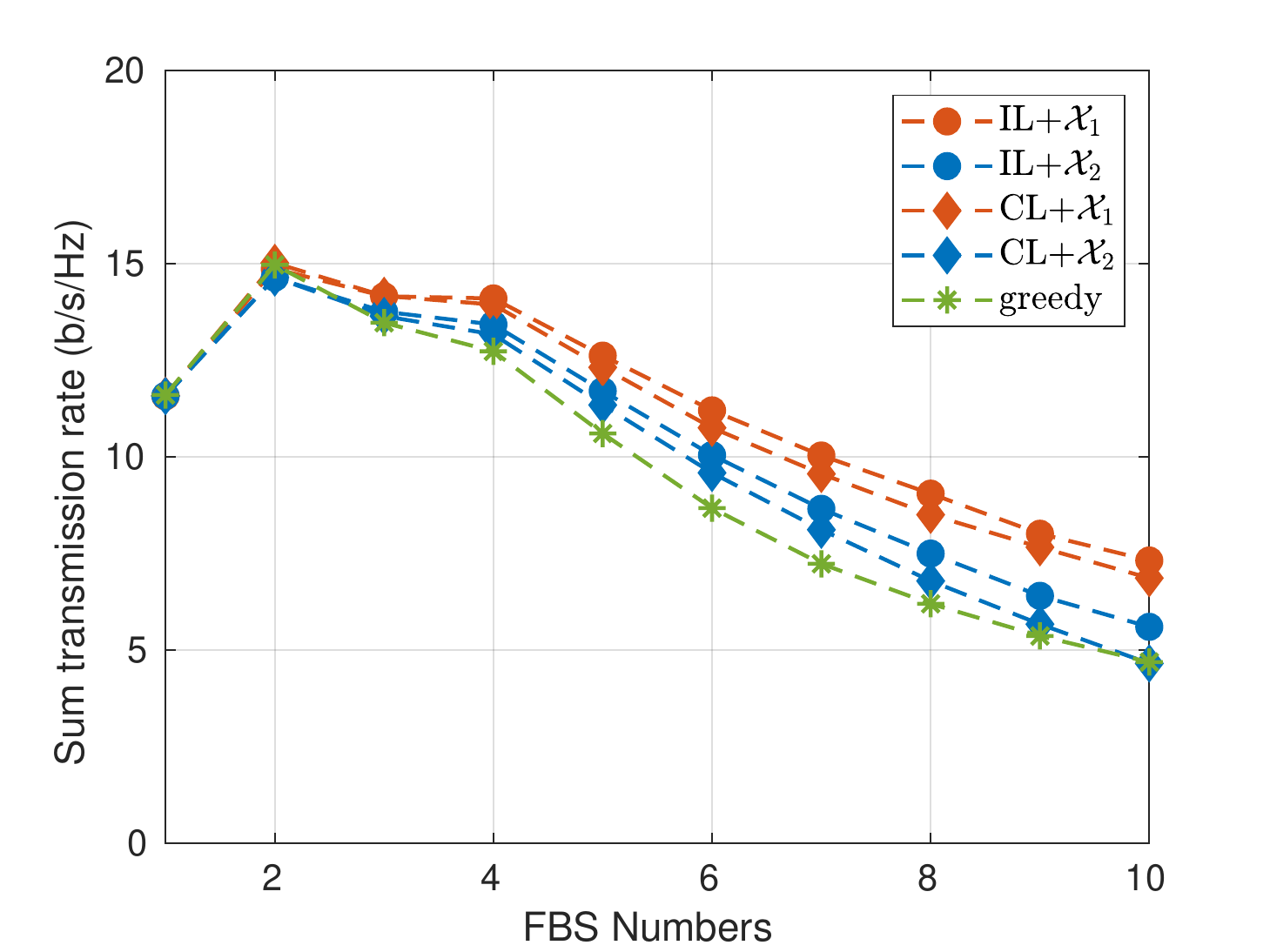}
        \caption{}\label{fig:sum}
    \end{subfigure}
    \begin{subfigure}[t]{.450\textwidth}
        \centering
        \includegraphics[width=1.0\columnwidth]{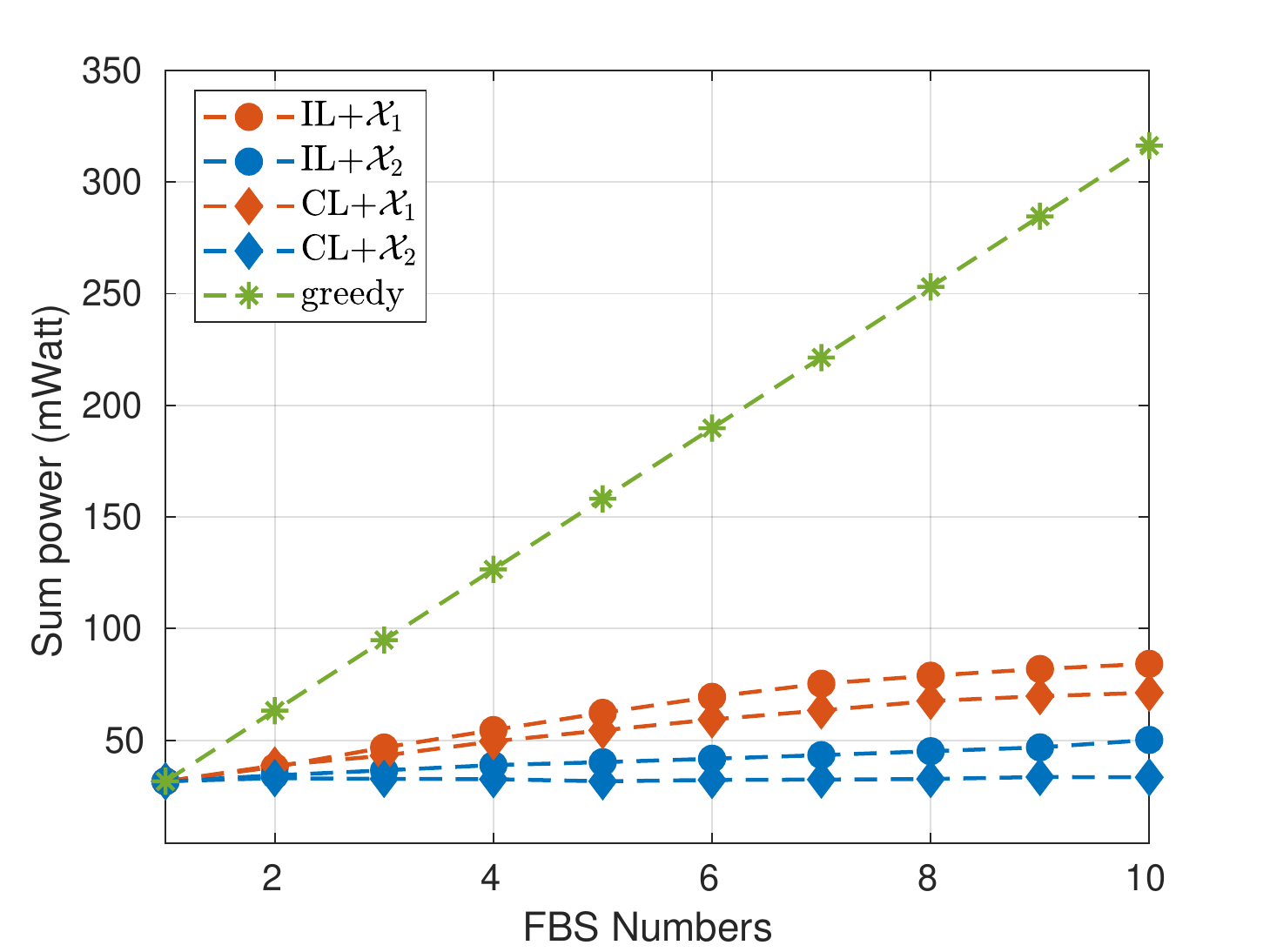}
        \caption{}\label{fig:power}
    \end{subfigure}%
    \caption{Performance of different learning configurations: (a) transmission rate of the MUE, (b) sum transmission rate of the FUEs, (c) sum transmit power of the FBSs.}
\end{figure}

Here, we show the simulation results of distributed power allocation with Q-DPA. First, we define two different state sets. The sets are defined as $\mathcal{X}_1=\left\lbrace X_1, X_3, X_4\right\rbrace$ and $\mathcal{X}_2=\left\lbrace X_2, X_3, X_4\right\rbrace$. In both sets, FBSs are aware of their relative location to the MUE and the MBS due to the presence of $X_3$ and $X_4$, respectively. The state set $\mathcal{X}_1$ gives knowledge of the status of the FUE to the FBS, and the state set $\mathcal{X}_2$ provides knowledge of the status of the MUE to the FBS. 

In order to understand the effect of independent and cooperative learning, and the effect of different state sets, we use four different learning configurations as: independent learning with each of the two state sets as IL+$\mathcal{X}_1$ and IL+$\mathcal{X}_2$, and cooperative learning with each of the two state sets as CL+$\mathcal{X}_1$ and CL+$\mathcal{X}_2$. The results are compared with \textit{greedy} approach in which each FBS chooses maximum transmit power. The simulation results are shown in three figures as: transmission rate of the MUE (Fig.~\ref{fig:MUE}), sum transmission rate of the FUEs (Fig.~\ref{fig:sum}), and sum transmit power of the FBSs (Fig.~\ref{fig:power}).
\begin{table}
\centering
\setlength\doublerulesep{1.0pt}
\caption{Performance of different learning configurations. $1$ is the best, and $4$ is the worst.}
\label{table_4}
\begin{tabular}{cccc} \hhline{====}
    \text{Learning configuration} & \text{$\sum p_k$}   &\text{$\sum r_k$}    &\text{$r_0$}   \\ \midrule
	IL+$\mathcal{X}_1$         & $4$  & $1$ & $4$\\ \midrule
	CL+$\mathcal{X}_1$         & $3$  & $3$ & $3$\\ \midrule
	IL+$\mathcal{X}_2$         & $2$  & $2$ & $2$\\ \midrule
	CL+$\mathcal{X}_2$         & $1$  & $4$ & $1$\\ \bottomrule  \\ 
\end{tabular}
\end{table}

According to Fig.~\ref{fig:power}, in the greedy algorithm, each FBS uses the maximum available power for transmission. Therefore, the greedy method introduces maximum interference for the MUE and has the lowest MUE transmission rate in Fig.~\ref{fig:MUE}. On the other hand, despite using maximum power, the greedy algorithm does not achieve highest transmission rate for the FUEs either (Fig.~\ref{fig:sum}). This is again due to the high level of interference.

The state set $\mathcal{X}_2$ provides knowledge of MUE's QoS status to the learning FBSs. Therefore, as we see in Fig.~\ref{fig:MUE}, the performance of IL with $\mathcal{X}_2$ is higher than the ones with $\mathcal{X}_1$. This statement is true for CL too. We can see the reverse of this conclusion in the FUEs' sum transmission rate in Fig.~\ref{fig:sum}. The performance of IL with $\mathcal{X}_1$ is higher than IL with $\mathcal{X}_2$. This is because the FBSs are aware of the status of the FUE, therefore, they consider actions that result in the state variable $X_1=\mathbbm{1}_{\left\lbrace \gamma_k \geq \Gamma_k \right\rbrace}$ to be $1$. This is true in comparison of the states in CL too. In conclusion, the state set $\mathcal{X}_1$ works in favor of femtocells and the state set $\mathcal{X}_2$ benefits the MUE.



We conclude from the simulation results that IL and CL present different trade-offs. More specifically, IL supports a higher sum transmission rate for the FBSs and a lower transmission rate for the MUE, while CL can support a higher transmission rate for the MUE at the cost of an overall lower sum transmission rate for the FBSs. From a power consumption point of view, IL results in a higher power consumption when compared to that of CL. In general, IL trains an FBS to be selfish compared to CL. IL can be very useful when there is no means of communication between the agents. On the other hand, CL trains an FBS to be more considerate about other FBSs at the cost of communication overhead.

In Table~\ref{table_4}, we have compared the performance of the four learning configurations. In each column, number $1$ is used as a metric to refer to the highest performance achieved and number $4$ is used to refer to the lowest performance observed. The first column represents the summation of transmit powers of FBSs, the second column indicates the summation of transmission rates of the FUEs, and the third column denotes the transmission rate of the MUE.

\subsection{Reward Function Performance}\label{sec_rewardPerformance}
\begin{figure}\label{fig:final}
\vspace*{-0.4cm}
    \centering
    \begin{subfigure}[t]{.450\textwidth}
        \centering
        \includegraphics[width=1.0\columnwidth]{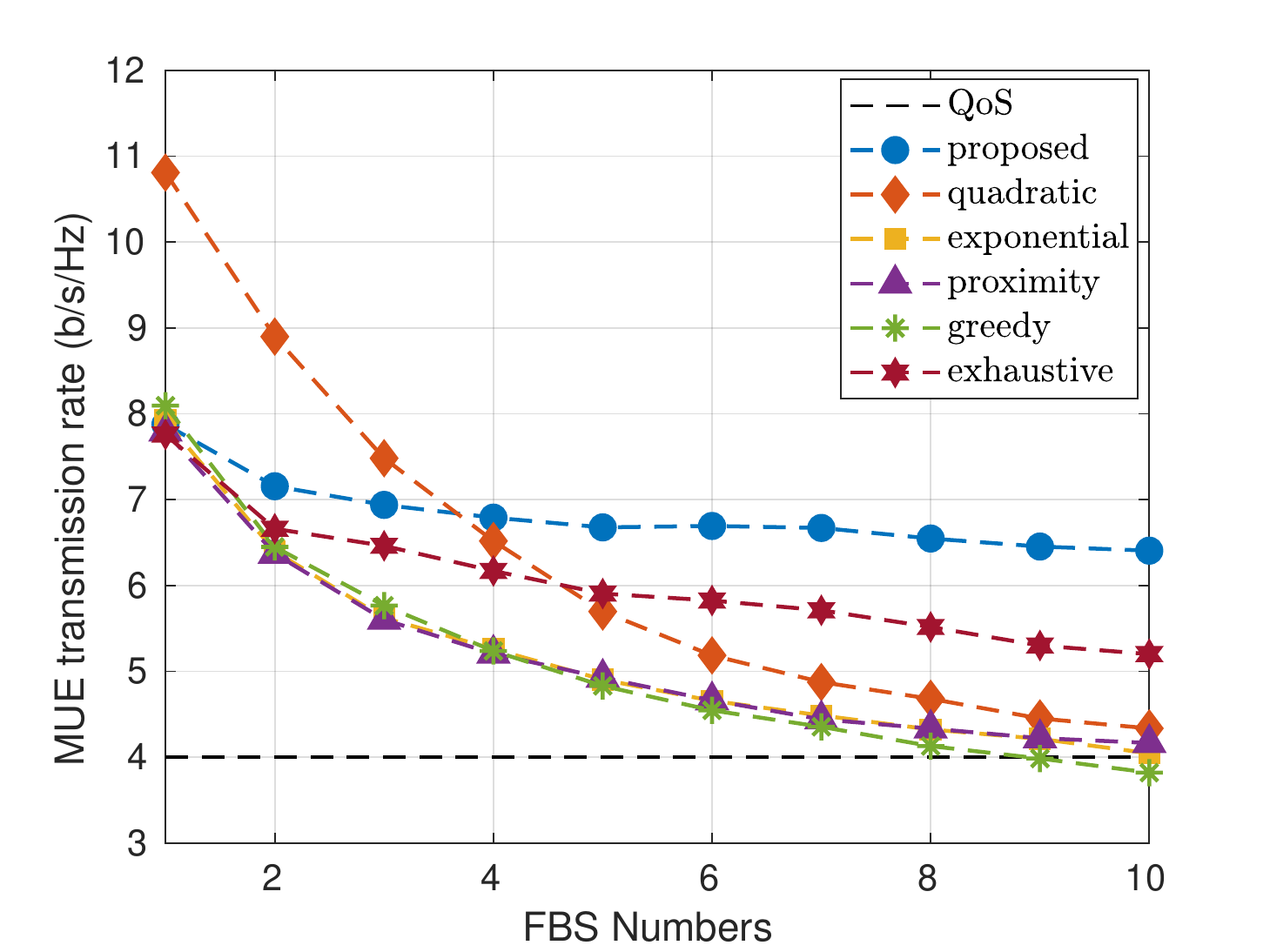}
        \caption{}\label{fig:MUE_1}
    \end{subfigure}%
    \begin{subfigure}[t]{.450\textwidth}
        \centering
        \includegraphics[width=1.0\columnwidth]{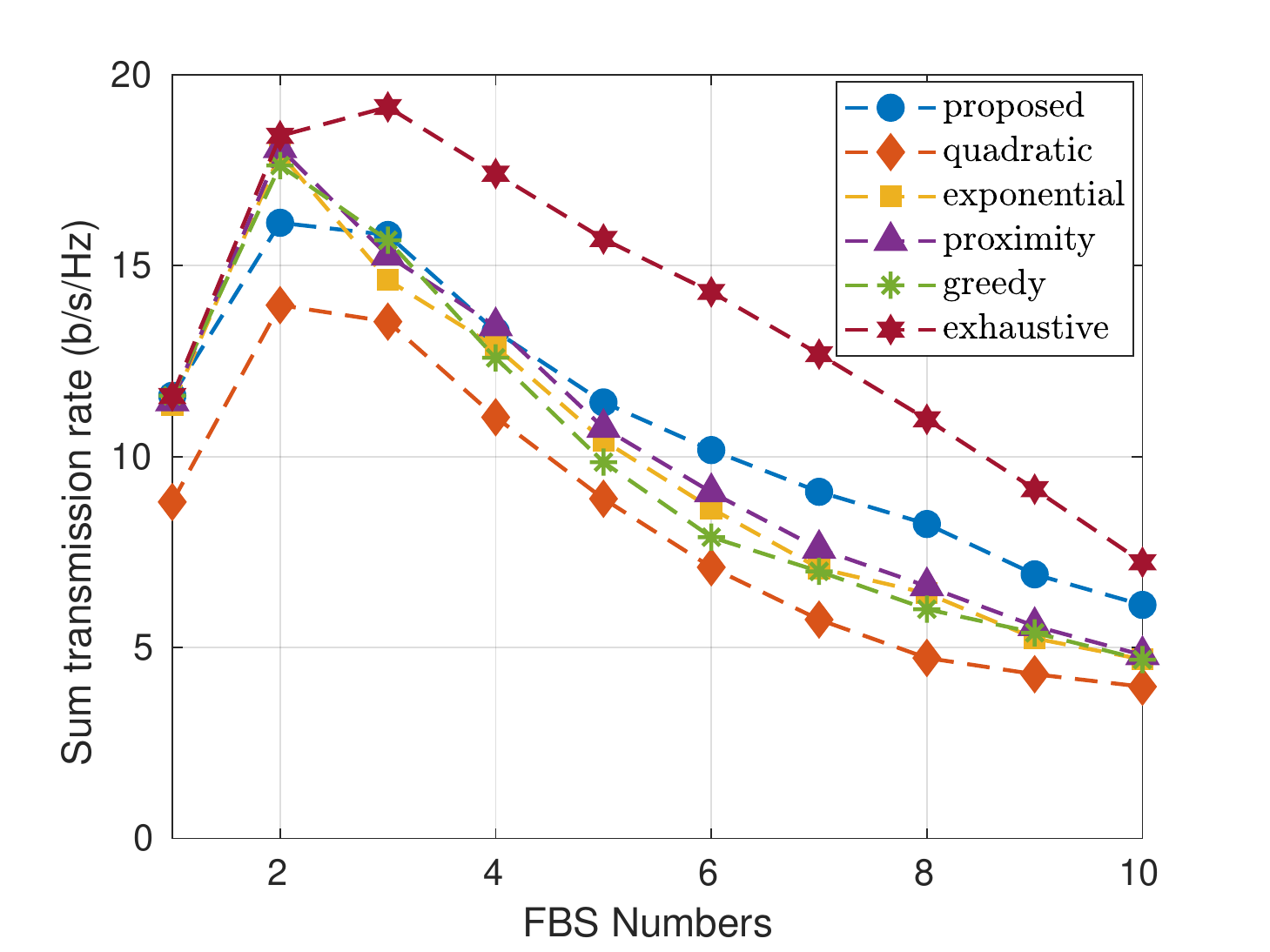}
        \caption{}\label{fig:sum_1}
    \end{subfigure}
    \begin{subfigure}[t]{.450\textwidth}
        \centering
        \includegraphics[width=1.0\columnwidth]{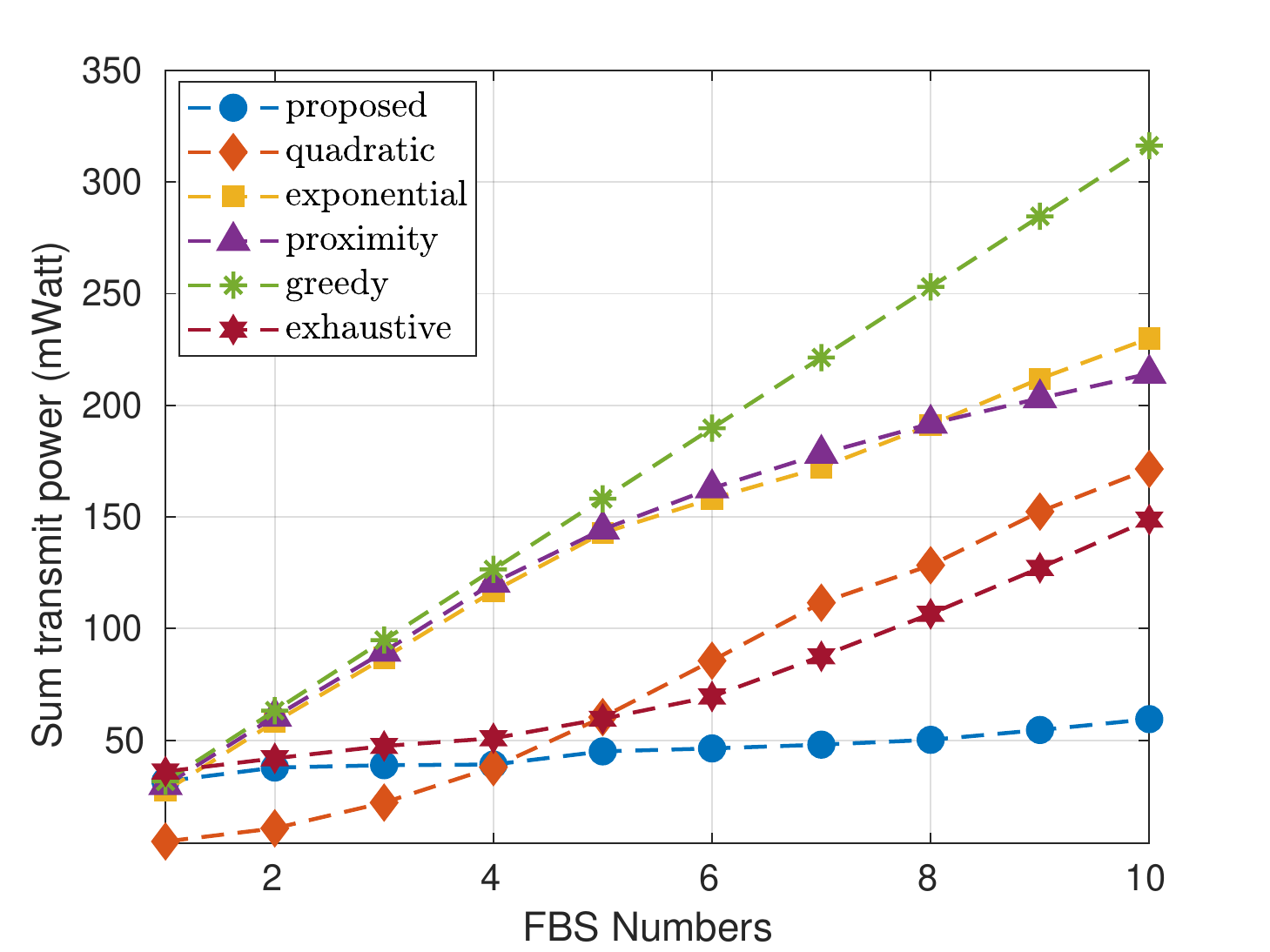}
        \caption{}\label{fig:power_1}
    \end{subfigure}%
    \caption{Performance of the proposed reward function compared to quadratic, exponential and proximity reward functions: (a) transmission rate of the MUE, (b) sum transmission rate of the FUEs, (c) sum transmit power of the FBSs.}
\end{figure}

Here, we compare the performance of the four reward functions discussed in Section~\ref{sec_reward_design}. Since the objective is to maximize the sum transmission rate of the FUEs, according to Table~\ref{table_4}, we choose the combination IL+$\mathcal{X}_1$ as the learning configuration. The performance of the reward functions are provided as the MUE transmission rate (Fig.~\ref{fig:MUE_1}), sum transmission rate of the FUEs (Fig.~\ref{fig:sum_1}), and sum transmission power of the FBSs (Fig.~\ref{fig:power_1}). In each figure, the solution of the optimization problem with exhaustive search and the performance of greedy method are provided. The exhaustive search provides us with the highest achievable sum transmission rate for the network. The quadratic, exponential, and proximity reward functions result in fast decaying of MUE transmission rate, while the proposed reward function results in a much slower decrease of the rate for the MUE. The proposed reward function manages to achieve a higher sum transmission rate compared to that of the other three reward functions as well. Fig.~\ref{fig:power_1} indicates that the proposed reward function reduces the sum transmitted power at the FBSs which in turn could result in lower levels of interference at the FUEs. In comparison with the exhaustive search solution as the optimal solution, there is a gap of performance. For instance according to Fig.~\ref{fig:power_1}, for eight number of FBSs, the proposed reward function uses an average of $50$~mWatt less sum transmit power than the optimal solution. However, as we see in Fig.~\ref{fig:sum_1} and Fig.~\ref{fig:MUE_1}, by using more power, the sum transmission rate can be improved and the transmission rate of the MUE can be decreased to the level of exhaustive solution without violating its minimum required rate. In our future works, we wish to cover this gap by using neural networks as the function approximator of the learning method.
\vspace{-20pt}
\section{Conclusion and Future Work} \label{sec_con}

In this paper, we propose a learning framework for a two-tier femtocell network. The framework enables addition of a new femtocell to the network, while the femtocell trains itself to adapt its transmit power to support its serving user while protecting the macrocell user. On the other hand, the proposed method as a distributed approach can solve the power optimization problem in dense HetNets, while significantly reducing power usage. The proposed framework is generic and motivates the design of machine learning based SONs for management schemes in femtocell networks. Besides, the framework can be used as a bench test for evaluating the performance of different learning configurations such as Markov state models, reward functions and learning rates. Further, the proposed framework can be applied to other interference-limited networks such as cognitive radio networks as well. 

In future work, it would be interesting to consider mmWave-enabled femtocells in the present setup. In fact, the high pathloss and shadowing along with the vulnerability of mmWave directional signals to the blockages impacts the learning outcome~\cite{art_slmz2}. This will in turn affect the subsequent power optimization problem. In addition, as we discussed in simulation section in details, there is a performance gap between the proposed approach and the exhaustive search. Although, the proposed approach results in less computational complexity; we wish to improve and cover this gap by utilizing neural networks as the function approximator of the learning method. In fact, neural networks can handle the large state-action spaces more efficiently. Moreover, another future complementary work to achieve a higher sum data rate and fill the performance gap would be to feed the interference model of the network to the factorization process. This way, a better factorization can be provided for the global Q-function.



\appendices
\newcommand{\bel}{\mathtt{T}{Q}} 
\newcommand{\belp}{\mathtt{T}{Q^\prime}} 
\newcommand{\belo}{\mathtt{T}{Q^*}}
\newcommand{\belk}{\mathtt{T}_k{Q_k}} 
\newcommand{\beli}{\mathtt{T}_i{Q_i}} 

\section{Proof of Proposition~\ref{theorem1}}\label{appendixTheorem1}
\begin{proof}
Assume an MDP represented as $\left(\mathcal{X}, \mathcal{A}, \Pr\left(y|x,a\right), r\left(x,a\right)\right)$, a policy $\pi$ with value-function $V_{\pi}:\mathcal{X} \rightarrow \mathbb{R}$ and Q-function $Q_{\pi}:\mathcal{Z} \rightarrow \mathbb{R}$, $\mathcal{Z} = \mathcal{X} \times \mathcal{A}$. Here, $\mathcal{A}$ refers to action space of one agent and $k$ is the iteration index. According to~\eqref{eq_value_func}, the maximum of the value-function can be fined as $V_{max}=\frac{R_{max}}{1-\beta}$. The Bellman optimality operator is defined as $\left(\bel\right)\left(x,a\right) \triangleq r\left(x,a\right) + \beta \sum_{y\in\mathcal{X}} \Pr\left(y|x,a\right) \underset{b \in \mathcal{A}}\max~Q\left(y,b\right)$. $\bel$ is a contraction operator with factor $\beta$, i.e., $\norm{\bel - \belp} \leq \beta \norm{Q-Q^\prime}$ and $Q^*$ is a unique fixed-point of $\left(\bel\right)\left(x,a\right)$, $\forall \left(x,a\right) \in \mathcal{Z}$. Further, for the ease of notation and readability the time step notation is slightly changed as $Q_k$ refers to the action-value function after $k$ iterations.

Assume that the state-action pair $\left(x,a\right)$ is visited $k$ times and $\mathcal{F}_k = \left\lbrace y_1, y_2, ..., y_{k}\right\rbrace$ are the visiting next states. At time step $k+1$, the update rule of Q-\textit{learning} is
\vspace{-10pt}
\begin{align}
Q_{k+1}\left(x,a\right) = \left(1-\alpha_k\right) Q_k\left(x,a\right) + \alpha_k \belk\left(x,a\right),
\end{align}
where, $\belk$ is the empirical Bellman operator defined as $\belk\left(x,a\right) \triangleq r\left(x,a\right) + \beta \underset{b \in \mathcal{A}}\max~Q\left(y_k,b\right)$. (From this point, for simplicity, we remove the dependency on $\left(x,a\right)$). It is easy to show that $E\left[\belk\right]=\bel_k$, therefore, we define $e_k$ as the estimation error of each iteration as $e_k= \belk-\bel_k$. By using $\alpha_k=\frac{1}{k+1}$, the update rule of Q-\textit{learning} can be written as
\vspace{-10pt}
\begin{align}\label{eq_q_new}
Q_{k+1}=\frac{1}{k+1}\left(k Q_k + \bel_k + e_k\right).
\end{align}
Now, in order to prove Proposition~\ref{theorem1}, we need to state the following lemmas.
\begin{lemma}\label{lem1}
For any $k \geq 1$
\vspace{-10pt}
\begin{align}\label{eq_lem1}
Q_{k}=\frac{1}{k} \sum_{i=0}^{k-1} \beli =\frac{1}{k} \left( \sum_{i=0}^{k-1} \bel_i + \sum_{i=0}^{k-1} e_i \right).
\end{align}
\end{lemma}
\begin{proof}
We prove this lemma by induction. The lemma holds for $k=1$ as $Q_1=\mathtt{T}_0{Q}_0=\mathtt{T}{Q}_0+e_0$.
We now show that if the result holds for $k$, then it also holds for $k+1$. From~\eqref{eq_q_new} we have
\begin{align*}
Q_{k+1} &= \frac{k}{k+1} Q_k + \frac{1}{k+1}\left(\bel_k + e_k\right)
        = \frac{k}{k+1} \frac{1}{k} \left( \sum_{i=0}^{k-1} \bel_i + \sum_{i=0}^{k-1} e_i \right) + \frac{1}{k+1}\left(\bel_k + e_k\right)&&&&\\
        &= \frac{1}{k+1} \left( \sum_{i=0}^{k} \bel_i + \sum_{i=0}^{k} e_i \right).
\end{align*}
Thus~\eqref{eq_lem1} holds for $k \geq 1$ by induction.
\end{proof}

\begin{lemma}\label{lem2}
Assume that initial action-value function, $Q_0$, is uniformly bounded by $V_{max}$. Then, for all $k \geq 1$ we have $\norm{Q_k} \leq V_{max}$ and $\norm{Q^*-Q_k} \leq 2 V_{max}$.
\end{lemma}
\begin{proof}
We first prove that $\norm{Q_k} \leq V_{max}$ by induction. The inequality holds for $k=1$ as
\vspace{-10pt}
\begin{align*}
\norm{Q_1} &=\norm{\mathtt{T}_0 Q_0}=\norm{r+\beta \max Q_0} \leq \norm{r} + \beta \norm{Q_0} \leq R_{max} + \beta V_{max} = V_{max}.
\end{align*}
Now, we assume that for $1 \leq i \leq k$, $\norm{Q_k} \leq V_{max}$ holds. First, $\norm{\belk}=\norm{r+\beta\max Q_k} \leq \norm{r}+\beta \norm{\max Q_k} \leq R_{max} + \beta V_{max} = V_{max}$. Second, from Lemma~\ref{lem1} we have
\vspace{-10pt}
\begin{align*}
\norm{Q_{k+1}}&=\frac{1}{k+1}\norm*{\sum_{i=0}^{k} \beli} \leq \frac{1}{k+1} \sum_{i=0}^{k} \norm*{\beli} \leq V_{max}.
\end{align*}
Therefore, the inequality holds for $k \geq 1$ by induction. Now the bound on $\norm{Q^*-Q_k}$ follows $\norm{Q^*-Q_k} \leq \norm{Q^*}+\norm{Q_k} \leq 2V_{max}$.
\end{proof}

\begin{lemma}\label{lem3}
Assume that initial action-value function, $Q_0$, is uniformly bounded by $V_{max}$, then, for any $k \geq 1$
\vspace{-10pt}
\begin{align}\label{eq_lem3}
\norm{Q^*-Q_k} \leq \frac{2\beta V_{max}}{k\left(1-\beta\right)} + \frac{1}{k} \norm*{\sum_{i=0}^{k-1} e_i}.
\end{align}
\end{lemma}
\begin{proof}
From Lemma~\ref{lem1}, we have
\vspace{-10pt}
\begin{align*}
Q^*-Q_k &= Q^* - \frac{1}{k} \left( \sum_{i=0}^{k-1} \bel_i + \sum_{i=0}^{k-1} e_i \right)
        = \frac{1}{k} \sum_{i=0}^{k-1}\left( \bel^*-\bel_i\right) - \frac{1}{k} \sum_{i=0}^{k-1} e_i.
\end{align*}
Therefore, we can write
\vspace{-10pt}
\begin{align*}
\norm{Q^*-Q_k} &\leq \frac{1}{k} \norm*{ \sum_{i=0}^{k-1}\left( \bel^*-\bel_i\right)} + \frac{1}{k} \norm*{\sum_{i=0}^{k-1} e_i}
               \leq \frac{1}{k} \sum_{i=0}^{k-1}\norm{  \bel^*-\bel_i} + \frac{1}{k} \norm*{\sum_{i=0}^{k-1} e_i}&&&&\\
               &\leq \frac{\beta}{k} \sum_{i=0}^{k-1}\norm{  Q^*-Q_i} + \frac{1}{k} \norm*{\sum_{i=0}^{k-1} e_i}.
\end{align*}
and according to~\cite{art_strehl}, $\norm{Q^*-Q_i} \leq \beta^i \norm{Q^*-Q_0}$. Hence, using Lemma~\ref{lem2}, we can write
\vspace{-10pt}
\begin{align*}
\norm{Q^*-Q_k} \leq \frac{\beta}{k} \sum_{i=0}^{k-1} 2 \beta^i V_{max} + \frac{1}{k} \norm*{\sum_{i=0}^{k-1} e_i}
               \leq \frac{2\beta V_{max}}{k\left(1-\beta\right)} + \frac{1}{k} \norm*{\sum_{i=0}^{k-1} e_i}.
\end{align*}
\end{proof}
Now, we prove Proposition~\ref{theorem1} by using the above result in Lemma~\ref{lem3}. To this aim, we need to provide a bound on the norm of the summation of errors in the inequality of Lemma~\ref{lem3}. First, we can write
\vspace{-10pt}
\begin{align*}
\frac{1}{k} \norm*{\sum_{i=0}^{k-1} e_i} &= \frac{1}{k} \underset{\left(x,a\right) \in \mathcal{Z}}\max \abs*{\sum_{i=0}^{k-1} e_i}.
\end{align*}
For the estimation error sequence $\left\lbrace e_0, e_1,\cdots, e_{k}\right\rbrace$, we have the property that $\mathbb{E} \left[ e_k| \mathcal{F}_{k-1}\right] =0 $ which means that the error sequence is a martingale difference sequence with respect to $\mathcal{F}_k$. Therefore, according to Hoeffding-Azuma inequality~\cite{art_hoeffding} for a martingale difference sequence of $\left\lbrace e_0, e_1,\cdots, e_{k-1}\right\rbrace$ which is bounded by $2V_{max}$, for any $t>0$, we can write 
\begin{align*}
\Pr\left(\abs*{\sum_{i=0}^{k-1} e_i} > t\right) \leq 2\exp\left(\frac{-t^2}{8kV^2_{max}}\right).
\end{align*}
Therefore, by a union bound over the state-action space, we have
\vspace{-10pt}
\begin{align*}
\Pr\left(\norm*{ \sum_{i=0}^{k-1} e_i} > t\right) \leq 2 \abs{\mathcal{X}}. \abs{\mathcal{A}}\exp\left(\frac{-t^2}{8kV^2_{max}}\right)=\delta,
\end{align*}
and then,
\vspace{-10pt}
\begin{align*}
\Pr\left(\frac{1}{k} \norm*{\sum_{i=0}^{k-1} e_i} \leq V_{max} \sqrt{\frac{8}{k}\ln{\frac{2\abs{\mathcal{X}}.\abs{\mathcal{A}}}{\delta}}}\right) \geq 1-\delta.
\end{align*}
Hence, with probability at least $1-\delta$ we can say
\vspace{-10pt}
\begin{align*}
\norm{Q^*-Q_k} \leq \frac{2 R_{max}}{\left(1-\beta\right)} \left[ \frac{\beta}{k\left(1-\beta\right)}+ \sqrt{\frac{2}{k}\ln{\frac{2\abs{\mathcal{X}}.\abs{\mathcal{A}}}{\delta}}} \right].
\end{align*}
Consequently, the result in Proposition~\ref{theorem1} is proved.
\end{proof}


%





\ifCLASSOPTIONcaptionsoff
  \newpage
\fi



\bibliographystyle{IEEEtran}
\bibliography{IEEEabrv,reference}

%








\end{document}